\documentclass[aps,prb,groupedaddress,twocolumn]{revtex4-1}

\usepackage{graphicx}
\usepackage{amsmath}
\usepackage{braket}
\usepackage{appendix}

\begin{document}


\title{Low-energy effective interactions beyond cRPA by the functional renormalization group}


\author{Michael Kinza}
\email[]{kinza@physik.rwth-aachen.de}
\author{Carsten Honerkamp}
\affiliation{
Institute for Solid State Theory, RWTH Aachen University,
D-52056 Aachen and JARA - Fundamentals of Future Information
Technology\\
}

\date{\today}

\begin{abstract}
In the derivation of low-energy effective models for solids targeting the bands near the Fermi level, the constrained random phase approximation (cRPA) has become an appreciated tool to compute the effective interactions. The  Wick-ordered constrained functional renormalization group (cfRG) generalizes the cRPA approach by including all interaction channels in an unbiased way. Here we present applications of the cfRG to two simple multi-band systems and compare the resulting effective interactions to the cRPA.  First we consider a multiband model for monolayer graphene, where we integrate out the $\sigma$-bands to get an effective theory for $\pi$-bands. It turns out that terms beyond cRPA are strongly suppressed by the different $xy$-plane reflection symmetry of the bands. In our model the cfRG-corrections to cRPA become visible when one disturbs this symmetry difference slightly, however without qualitative changes. This study shows that the embedding or layering of two-dimensional electronic systems can alter the effective interaction parameters beyond what is expected from screening considerations.
The second example is a one-dimensional model for a diatomic system reminiscent of a CuO chain, where we consider an effective theory for Cu 3d-like orbitals. Here the fRG data shows relevant and qualitative corrections compared to the cRPA results. We argue that the new interaction terms affect the magnetic properties of the low-energy model.
\end{abstract}


\maketitle

\setlength{\parindent}{0pt}
\section{Introduction}
The study of correlation and ordering effects in solids on lower energy scales is very often performed by using low-energy effective models (LEEM). In formulating such a model, much of the material-specific information has to be cast into the parameters of the LEEM. A standard example for this approach is the one-band Hubbard model. In its derivation (see e.g. Ref. \onlinecite{Au94}), a single band at the Fermi level, held responsible for correlation effects like magnetism and unconventional superconductivity, is chosen as single-particle basis for the LEEM. The hopping parameters of the Hubbard model encode the material-specific extent and energetics of Wannier states arising from the electronic and crystal structure. Similarly, the interaction parameter of the Hubbard model, $U$, depends on the electronic structure via Coulomb integrals of the Wannier basis, and additional screening by other processes, as will be detailed further below. Once the Hubbard model is formulated, many-body methods capable of capturing electronic correlations to sufficient detail can be used to solve for the low-energy properties. Usually, this last step is already a complicated endeavor, and for a long time, most researchers have focussed on getting the qualitative features of the model physics right, e.g. to understand the Mott transition (see, e.g. Ref. \onlinecite{Geo96}). In recent time however, quantitative aspects have attracted more interest again. The combination of ab-initio methods with dynamical mean-field theory has become a lively field of activities (see, e.g. Ref. \onlinecite{Kot04,Pav11}).
In the field of high-temperature superconductivity, various groups have tried to understand the systematics of the $T_c$-differences between various high-$T_c$ cuprates \cite{Ken08,Sak10,Sak12,Sak14} or iron arsenides.\cite{Pla11,Lie14} Likewise, in the field of graphene, an understanding of the ground state of a single graphene sheet as well as of its layered cousins requires a more precise understanding what the actual model parameters are.\cite{Hon08,Sche12,Sche12b}

Regarding the effective model parameters, there may be a different degree of controlled knowledge about single-particle parameters, like hopping amplitudes, compared to what is usually known about the interaction parameters. The single-particle parameters can, at least in a dressed form, be estimated to large extent from experiments like angle-resolved photoemission. De-Haas-van-Alphen and angular-dependent magneto-resistance oscillation techniques are additional established complementary tools to study the renormalized single-particle properties. In contrast with this, for the interaction parameters, there are at best indirect probes to their renormalized values, and even then the correspondence to parameters of a chosen LEEM may involve additional theoretical steps. Therefore, this part of the LEEM description is much less certain, may however be of crucial importance to understand material-specific interaction effects in quantitative form.  Fortunately, there have been at least two useful theoretical approaches that try to compute the relevant interaction parameters starting from the ab-initio electronic structure, beyond the direct calculation of Coulomb integrals in a localized basis set. These are the constrained local density approximation,\cite{Gun89,Ani91} and the recently strongly used constrained random phase approximation (cRPA, \cite{Ary04,Miy08,Miy09}). In the latter approach, the Coulomb integrals in the localized basis are altered by dielectric functions arising the additional screening due to the bands not included in the LEEM. This screening is obtained from computing the particle-hole polarizability of the band structure constrained to those processes that do not happen exclusively within the target bands of the LEEM, as those will be dealt with in the solution of the LEEM and should not be treated twice. As is clear from the name cRPA, this approach is an approximation with caveats that are shared by many other RPA approaches. It is clear however that the cRPA captures important physical ingredients and it has been a very welcomed innovation over the last years. 

Here we ask the question if one can improve on the cRPA in a systematic way, in order to get even more quantitative precision. From other instances in physics where RPA and related perturbation expansions are used, it is known that RPA in the charge channel with a polarizability as result can overlook important parts of physics. Moreover, it is not clear to us which small parameter warrants that an RPA summation for the LEEM interactions is necessarily close to the full answer.  In fact, that perturbation channels different from the particle-hole RPA terms lead to important renormalizations has been well known in condensed matter physics for a long time. Anderson and Morel \cite{Brue60,Mor62} showed that the particle-particle channel causes a strong suppression of the electron-electron repulsion at energy scales above the phonon frequencies, opening the chance that weak phonon-induced interactions lead to Cooper pairing. Around the same time, Kanamori \cite{Kan63} employed formally related calculations again exploiting the particle-particle channel in order to compute effective interactions for magnetic ordering. Hence, it is not surprising that the particle-particle channel will also affect the low-energy effective interactions in a multi band system. A quantitative difference may be that in the multiband case treated in this paper, the 'lower cutoff' at which the renormalizations of interest are computed is still of the order of electron volts, and not much below as, e.g., in the Cooper pairing problem. Hence the logarithmic contributions in the particle-particle channel do not come to full strength, and the corrections are not as strong. Yet, as we will show below, they can be significant. Moreover similar things with noticeable magnitude happen in the 'magnetic' crossed particle-hole as well.

In an earlier publication, one of us laid out a {\em constrained  functional renormalization group} (cfRG) formalism that goes beyond the cRPA by including the other channels mentioned above plus vertex corrections, but can be reduced to cRPA by dropping all but one one-loop diagram. 
In this earlier work, only very simplistic models were considered to show that the more general scheme can indeed lead to different answers. Here we use the cfRG in two more realistic model settings in order to show under which circumstances differences in the LEEM interactions can occur. The two models are {\em 1)} a simple 8-band model for a graphene sheet, that in addition to the two $\pi$-bonding orbitals per unit cell that give rise to Dirac spectrum also contains the six $sp^2$-$\sigma$-bonding orbitals per unit cell; and {\em 2)} an one-dimensional $dps$-model that mimics copper oxides or similar materials. Here the hopping between $d$-orbitals on Cu sites goes through $p$-orbitals on O sites. 
The main findings of this work are that in model {\em  1)} the cRPA gives a rather faithful picture of the LEEM interaction. Here the different symmetry of targets ($\pi$-) bands and the ($\sigma$-) bands that are integrated out is the key factor for the absence of major corrections. Visible albeit limited corrections can however be provoked by breaking the symmetry of the graphene plane with respect to inversion of the perpendicular coordinate, i.e. by a substrate or a perpendicular electric field. 
 In model {\em 2)} no such symmetry protection of the cRPA exists and we find rather marked deviations in the cFRG. 
Notably, the pairing and magnetic particle-hole channel cause distinct new features in the LEEM interactions, and the down-screening of the density-density terms is actually weakened compared to the cRPA result. We also try to understand what physical consequences arise from the extra terms in the effective interaction. Here, the main finding is an enhanced tendency of the LEEM to develop a large-spin ground state. This shows that the LEEM beyond the cRPA might even lead to new qualitative physical effects.

This paper is organized in the following way: In Sec.\ref{formalism} we give a brief introduction to the constrained fRG method and discuss how the frequency structure of the vertex function can be interpretated in terms of density-density and spin-spin interactions.
In Sec. \ref{results} we describe our results for the graphene and the Cu-O-chain model. Sec.\ref{conclusions} contains the conclusions and a brief outlook.

\section{Formalism}\label{formalism}
Here we review the Wick-ordered fRG formalism that, as argued in Ref. \onlinecite{Hon12}, can be used to compete the LEEM interactions beyond cRPA.
\subsection{Wick-ordered fRG}
The basic setup of what follows is an electronic structure that can be divided in two sets of bands. The goal is to integrate out one set of bands (called 'high-energy bands', although an energetic hierarchy is not essential for the formalism) in order to compute an effective description for the remaining bands that are usually close to the Fermi level and that shall be called 'target bands'.

In order to integrate out the high-energy bands we use the Wick-ordered fRG approach, described in Ref. \onlinecite{Hon12}, which lives in the functional integral formalism that describes the band electrons as anti-commuting Grassmann fields.
The quantum numbers carried by these fields are then a fermionic Matsubara frequency $i\omega_n$, a wave vector $\vec{k}$ in the first Brillouin zone of the lattice, the
band-index $b$ and a spin projection $s \in \lbrace\uparrow $, $\downarrow\rbrace$. The band energies are then given by $\epsilon_b (\vec{k})$. We assume spin rotational invariance, i.e. the band are spin-degenerate. 
To parameterize the two-particle vertex function (ie. 'the interaction') we use bosonic 'Mandelstam' Matsubara frequencies, which are related to the fermionic frequencies of the incoming (1 and 2) and outgoing (1' and 2') particles 
by the total incoming frequency $i\nu_1 = i\omega_1 + i\omega_2$, and the two transfers $i\nu_2 = i\omega_{1'}-i\omega_1$ and $i\nu_3 = i\omega_2 - i\omega_{1'}$. The  SU(2)-spin-rotational invariance is reflected in our notation for the vertex function by the requirement that particle 1 and 1' have the same spin projection $s_1=\pm \frac{1}{2}$, and 2 and 2' the same $s_2=\pm \frac{1}{2}$.

The fRG flow has the goal of integrating out the high-energy bands. It is organized in terms of Wick-ordered correlation functions\cite{Sal97,Sal99} and employs a generalized cutoff $\Lambda$ that, upon evolving from 1 to 0, removes the modes of the high-energy band from the theory and incorporates their effect in the remaining low-energy theory.\cite{Hon12} Compared to the usual amputated correlation functions that one obtains in the effective action by integrating out high-energy bands straightforwardly, the Wick-ordered two-particle correlation function is used as interaction of the LEEM. The Wick-ordered correlation function of $n$-th order contains in addition contributions from tree diagrams with $n+2m$ external legs built with high-energy propagators and one-particle-irreducible (1PI) vertex functions, where $2m$ legs have been contracted with $m$ low-energy propagators. This resummation is crucial for incorporating the 'mixed' loop diagrams with one line in the high-energy and the second line in the low-energy sector. For more concrete formulae regarding the connection between the different vertex functions, see the paper \onlinecite{Hon12} or textbooks like \onlinecite{Sal99,Kop10b}.

\begin{figure}[htbp]
    \centering
    \includegraphics[width=0.50\textwidth]{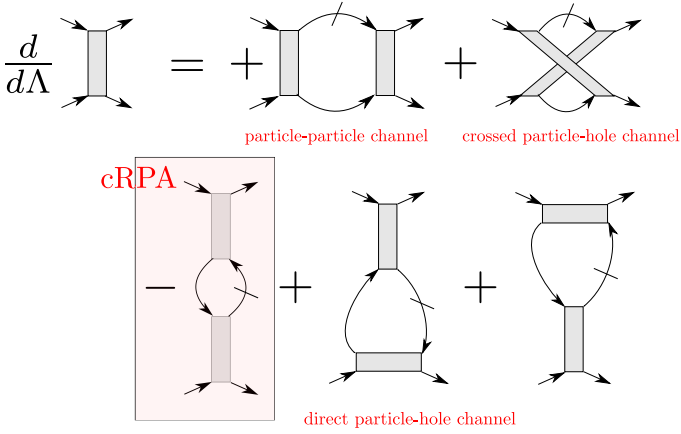}
    \caption[flowequations]{(color online) Diagrammatic representation of the fRG flow equations, in the truncation that drops the six-point vertex. 
    The dashed line carries a differentiated cutoff function and is only nonzero in the high-energy bands. The second loop line is a free propagator that can be in both high-energy and low-energy bands. Restricting the flow to the first diagram of the direct  particle-hole channel in the second row reproduces the cRPA approximation.} 
    \label{pictflowequations}
\end{figure}

For the integration over the modes of the high-energy bands the flow equations for the Wick-ordered vertices can be written down. As we focus in this work on the effective interactions, we will neglect self-energy corrections, although they may as well be of importance. Then fRG flow equations 
for the Wick-ordered two-particle interaction or coupling function $V^{\Lambda}(i\nu_1,i\nu_2,i\nu_3;\vec{k}_{1'},\vec{k}_{2'},\vec{k}_{1};b_{1'},b_{2'},b_{1},b_{2})$ 
are given by \cite{Sal97,Sal99}
\begin{align}
\frac{d}{d\Lambda}V^{\Lambda}=\dot\Phi^{\Lambda}_{\text{pp}}+\dot\Phi^{\Lambda}_{\text{dph}}+\dot\Phi^{\Lambda}_{\text{crph}}
\end{align}
with
\begin{widetext}
\begin{align}
\nonumber
\dot\Phi^{\Lambda}_{\text{pp}}(i\nu_1,i\nu_2,i\nu_3;\vec{k}_{1'},\vec{k}_{2'},\vec{k}_1;b_{1'},b_{2'},b_{1},b_{2})=&\frac{1}{\beta}\sum_{i\omega}\sum_{b,b',\vec{k}}\frac{\partial}{\partial\Lambda}\left[D^{\Lambda}(i\omega,\vec{k},b)D^{\Lambda}(i\nu_1-i\omega,\vec{k}_{1'}+\vec{k}_{2'}-\vec{k},b')\right]&\\
\nonumber
&\times V^{\Lambda}(i\nu_1,i\nu_{++-}-i\omega,i\nu_{+-+}-i\omega;\vec{k}_{1'},\vec{k}_{2'},\vec{k};b_{1'},b_{2'},b,b')&\\
\label{eqfloweq1}
&\times V^{\Lambda}(i\nu_1,i\omega-i\nu_{+--},i\nu_{+-+}-i\omega;\vec{k},\vec{k}_{1'}+\vec{k}_{2'}-\vec{k},\vec{k}_1;b,b',b_1,b_2)&\\
\nonumber
\dot\Phi^{\Lambda}_{\text{dph}}(i\nu_1,i\nu_2,i\nu_3;\vec{k}_{1'},\vec{k}_{2'},\vec{k}_1;b_{1'},b_{2'},b_{1},b_{2})=&\frac{1}{\beta}\sum_{i\omega}\sum_{b,b',\vec{k}}\frac{\partial}{\partial\Lambda}\left[D^{\Lambda}(i\omega,\vec{k},b)D^{\Lambda}(i\nu_2+i\omega,\vec{k}_{1'}-\vec{k}_1+\vec{k},b')\right]&\\
\nonumber
\big[&\underline{-2 V^{\Lambda}(i\nu_{++-}+i\omega,i\nu_2,i\omega-i\nu_{+--};\vec{k}_{1'},\vec{k},\vec{k}_{1};b_{1'},b,b_1,b')}&\\
\nonumber
&\underline{\times V^{\Lambda}(i\omega+i\nu_{+++},i\nu_2,i\nu_{+-+}-i\omega;\vec{k}_{1'}-\vec{k}_1+\vec{k},\vec{k}_{2'},\vec{k};b',b_{2'},b,b_2)}&\\
\nonumber
&+V^{\Lambda}(i\nu_{++-}+i\omega,i\nu_2,i\omega-i\nu_{+--};\vec{k}_{1'},\vec{k},\vec{k}_1;b_{1'},b,b_1,b')&\\
\nonumber
&\times V^{\Lambda}(i\omega+i\nu_{+++},i\omega-i\nu_{+-+},-i\nu_2;\vec{k}_{1'}-\vec{k}_1+\vec{k},\vec{k}_{2'},\vec{k}_2;b',b_{2'},b_2,b)&\\
\nonumber
&+V^{\Lambda}(i\omega+i\nu_{++-},i\omega-i\nu_{+--},i\nu_2;\vec{k},\vec{k}_{1'},\vec{k}_1;b,b_{1'},b_1,b')&\\
&\times V^{\Lambda}(i\omega+i\nu_{+++},i\nu_2,i\nu_{+-+}-i\omega;\vec{k}_{1'}-\vec{k}_1+\vec{k},\vec{k}_{2'},\vec{k};b',b_{2'},b,b_2)\big]&\\
\nonumber
\dot\Phi^{\Lambda}_{\text{crph}}(i\nu_1,i\nu_2,i\nu_3;\vec{k}_{1'},\vec{k}_{2'},\vec{k}_1;b_{1'},b_{2'},b_{1},b_{2})=&\frac{1}{\beta}\sum_{i\omega}\sum_{b,b',\vec{k}}\frac{\partial}{\partial\Lambda}\left[D^{\Lambda}(i\omega,\vec{k},b)D^{\Lambda}(i\nu_3+i\omega,\vec{k}+\vec{k}_{2'}-\vec{k}_1,b')\right]&\\
\nonumber
&\times V^{\Lambda}(i\omega+i\nu_{+-+},i\omega-i\nu_{+--},i\nu_3;\vec{k},\vec{k}_{2'},\vec{k}_1;b,b_{2'},b_1,b')&\\
\label{eqfloweq2}
&\times V^{\Lambda}(i\omega+i\nu_{+++},i\nu_{+--}-i\omega,i\nu_3;\vec{k}_{1'},\vec{k}+\vec{k}_{2'}-\vec{k}_1;b_{1'},b',b,b_2)&
\end{align}
\end{widetext}
Here we have defined $i\nu_{+-+}=\frac{1}{2}(i\nu_1-i\nu_2+i\nu_3)$ ($i\nu_{+++}$, $i\nu_{++-}$, $i\nu_{+--}$ are defined in an analogue way). 
Restricting the flow to the underlined part of the direct particle-hole channel is equivalent to the cRPA solution.\cite{Hon12} In Fig. \ref{pictflowequations} we show a diagrammatic representation of the flow equations  (\ref{eqfloweq1})-(\ref{eqfloweq2}).

The propagator $D^{\Lambda}$ is given by
\begin{align}
D^{\Lambda}(b)=\begin{cases}
(1-\Lambda)\mathcal{G}_{0} &\text{if}\ b\in \text{high energy bands} \\
\mathcal{G}_{0} &\text{if}\ b\in \text{low energy bands}
\end{cases}
\end{align}

As mentioned above, the flow equations are integrated from $\Lambda=0$ to $\Lambda=1$, in order that in the end of the flow the high energy bands are integrated out. Here two differences to usual momentum-shell RG schemes should be noted: First, we use a 'flat' cutoff that simply switches off the high-energy bands in the free part of action irrespective of their precise energy, and second, as characteristic feature of the Wick-flow, the second line in the one-loop terms is a low-energy propagator complementary to what has already been integrated out. Using a flat cutoff simplifies the numerical treatment significantly as long as no self-energy corrections are included. Then the diagrams on the right hand have to be commuted only once, and just get multiplied by $-(1-\Lambda)$ during the flow.

It is actually possible to treat  an interaction vertex that depends on three frequencies (see, e.g., Ref. \onlinecite{Ueb12}), but this leads to a larger numerical effort. In this study we choose to simplify the flow equations in the frequency domain and apply a Karrasch-Husemann decomposition.\cite{Hus09,Kar08} Here, the change of the interaction with $\Lambda$ is written as a sum of three functions that each depend on only one of the the three bosonic frequencies mentioned before, while the other two frequencies are kept at zero in these functions.  Then the full vertex that still depends on three frequencies but can be written as a sum of three terms that each depend on one frequency.
We conclude that the decomposition certainly suffices to study effects beyond RPA, but that at stronger couplings, even more frequency structures might emerge.\cite{Roh12,Kin13b}
In any case, this approximation leads to the simplified flow equations

\begin{widetext}
\begin{align}
\nonumber
\dot\Phi^{\Lambda}_{\text{pp}}(i\nu_1;\vec{k}_{1'},\vec{k}_{2'},\vec{k}_1;b_{1'},b_{2'},b_{1},b_{2})=&\frac{1}{\beta}\sum_{i\omega}\sum_{b,b',\vec{k}}\frac{\partial}{\partial\Lambda}\left[D^{\Lambda}(i\omega,\vec{k},b)D^{\Lambda}(i\nu_1-i\omega,\vec{k}_{1'}+\vec{k}_{2'}-\vec{k},b')\right]&\\
\nonumber
&\times V^{\Lambda}(i\nu_1,i\nu_{1}/2-i\omega,i\nu_{1}/2-i\omega;\vec{k}_{1'},\vec{k}_{2'},\vec{k};b_{1'},b_{2'},b,b')&\\
&\times V^{\Lambda}(i\nu_1,i\omega-i\nu_{1}/2,i\nu_{1}/2-i\omega;\vec{k},\vec{k}_{1'}+\vec{k}_{2'}-\vec{k},\vec{k}_1;b,b',b_1,b_2)&\\
\nonumber
\dot\Phi^{\Lambda}_{\text{dph}}(i\nu_2;\vec{k}_{1'},\vec{k}_{2'},\vec{k}_1;b_{1'},b_{2'},b_{1},b_{2})=&\frac{1}{\beta}\sum_{i\omega}\sum_{b,b',\vec{k}}\frac{\partial}{\partial\Lambda}\left[D^{\Lambda}(i\omega,\vec{k},b)D^{\Lambda}(i\nu_2+i\omega,\vec{k}_{1'}-\vec{k}_1+\vec{k},b')\right]&\\
\nonumber
\big[&-2 V^{\Lambda}(i\nu_{2}/2+i\omega,i\nu_{2},i\omega+i\nu_{2}/2;\vec{k}_{1'},\vec{k},\vec{k}_{1};b_{1'},b,b_1,b')&\\
\nonumber
&\times V^{\Lambda}(i\omega+i\nu_{2}/2,i\nu_2,-i\nu_{2}/2-i\omega;\vec{k}_{1'}-\vec{k}_1+\vec{k},\vec{k}_{2'},\vec{k};b',b_{2'},b,b_2)&\\
\nonumber
&+V^{\Lambda}(i\nu_{2}/2+i\omega,i\nu_2,i\omega+i\nu_{2}/2;\vec{k}_{1'},\vec{k},\vec{k}_1;b_{1'},b,b_1,b')&\\
\nonumber
&\times V^{\Lambda}(i\omega+i\nu_{2}/2,i\omega+i\nu_{2}/2,-i\nu_2;\vec{k}_{1'}-\vec{k}_1+\vec{k},\vec{k}_{2'},\vec{k}_2;b',b_{2'},b_2,b)&\\
\nonumber
&+V^{\Lambda}(i\omega+i\nu_{2}/2,i\omega+i\nu_{2}/2,i\nu_2;\vec{k},\vec{k}_{1'},\vec{k}_1;b,b_{1'},b_1,b')&\\
&\times V^{\Lambda}(i\omega+i\nu_{2}/2,i\nu_2,-i\nu_{2}/2-i\omega;\vec{k}_{1'}-\vec{k}_1+\vec{k},\vec{k}_{2'},\vec{k};b',b_{2'},b,b_2)\big]&\\
\nonumber
\dot\Phi^{\Lambda}_{\text{crph}}(i\nu_3;\vec{k}_{1'},\vec{k}_{2'},\vec{k}_1;b_{1'},b_{2'},b_{1},b_{2})=&\frac{1}{\beta}\sum_{i\omega}\sum_{b,b',\vec{k}}\frac{\partial}{\partial\Lambda}\left[D^{\Lambda}(i\omega,\vec{k},b)D^{\Lambda}(i\nu_3+i\omega,\vec{k}+\vec{k}_{2'}-\vec{k}_1,b')\right]&\\
\nonumber
&\times V^{\Lambda}(i\omega+i\nu_{3}/2,i\omega+i\nu_{3}/2,i\nu_3;\vec{k},\vec{k}_{2'},\vec{k}_1;b,b_{2'},b_1,b')&\\
&\times V^{\Lambda}(i\omega+i\nu_{3}/2,-i\nu_{3}/2-i\omega,i\nu_3;\vec{k}_{1'},\vec{k}+\vec{k}_{2'}-\vec{k}_1;b_{1'},b',b,b_2)&
\end{align}
\end{widetext}
In this approximation, we can also write the full coupling function in the target bands at the end of the flow as follows,
\begin{eqnarray} \label{Vdecomp}
\nonumber
&& V^{\Lambda=1}(i\nu_1,i\nu_2,i\nu_3;\vec{k}_{1'},\vec{k}_{2'},\vec{k}_{1};b_{1'},b_{2'},b_{1},b_{2})\\ 
\nonumber
&&= V_{\mathrm{target-bands}}  \\
\nonumber
&& + \Phi^{\Lambda=1}_{\text{pp}}(i\nu_1;\vec{k}_{1'},\vec{k}_{2'},\vec{k}_1;b_{1'},b_{2'},b_{1},b_{2})  \\
\nonumber
&& + \Phi^{\Lambda=1}_{\text{dph}}(i\nu_2;\vec{k}_{1'},\vec{k}_{2'},\vec{k}_1;b_{1'},b_{2'},b_{1},b_{2}) \\
&& + \Phi^{\Lambda=1}_{\text{crph}}(i\nu_3;\vec{k}_{1'},\vec{k}_{2'},\vec{k}_1;b_{1'},b_{2'},b_{1},b_{2}) \, .
\end{eqnarray} 
In the cRPA, the second and the fourth term on the right hand side are absent. Note that otherwise the flow for any $\Lambda$ couples the different contributions $\Phi^{\Lambda}_{\text{pp}}$, $\Phi^{\Lambda}_{\text{dph}}$ and $\Phi^{\Lambda}_{\text{crph}}$.

\subsection{Frequency structure of the vertex function} \label{vertex2hamilton}

As can be seen from the flow equations, the resulting vertex will depend on three frequencies. Unless one can insert this function directly in a many-body low-energy solver like CT-QMC, this rather rich frequency structure must be interpreted in some way. We propose to do this by separating the different channels and interpreting these as charge, spin and pairing terms of the effective interaction.  In the following we explain how this decomposition is organized. This is done most easily in real space on the lattice of the effective low-energy model. As for the problems studied explicitly in the remainder of the paper, the effective model with contain only one orbital per site, we can henceforth ignore the band or orbital indices in the notation.

Let us describe the fermions in the effective model with Grassmann fields  $c^{\dag}_{i,\sigma}(\tau) $ and $c_{i,\sigma} (\tau)$ in imaginary time. Then the Heisenberg density operator at site $i$ is given by
\begin{align}
\rho_i(\tau) &= \sum_{\sigma} c^{\dag}_{i,\sigma}(\tau) c_{i,\sigma} (\tau) \, , &\\
\rho_i(i\nu) &= \int_0^{\beta}d\tau \exp(i\nu\tau)\rho_i(\tau) = \frac{1}{\beta}\sum_{\sigma,i\omega}c^{\dag}_{i\sigma}(i\omega)c_{i\sigma}(i\omega+i\nu) \, .
\end{align}

A retarded density-density interaction term between site pairs $(i,j)$ (symmetric in these indices, and each pair $(i,j)$ counted once) can then be written as
\begin{widetext}
\begin{align}
S_{\text{charge}} &= \frac{1}{\beta}Ê\sum_{i\nu, (i,j) } f_{\text{charge}} (i\nu ,i,j ) \rho_i(i\nu)\rho_j(-i\nu)  \label{vrhorho}\\
&= \frac{1}{\beta^3}Ê\sum_{i\nu,  i,j, i\omega , i\omega' \atop \sigma, \sigma'  } f_{\text{charge}} (i\nu ,i,j ) 
c^{\dag}_{i\sigma}(i\omega)c^{\dag}_{j \sigma'}(i\omega') c_{j\sigma'}(i\omega'-i\nu) c_{i\sigma}(i\omega+i\nu) 
 \\
&=\frac{1}{\beta^3}\sum_{i\omega_{1'},i\omega_{2'}\atop i\omega_1,i\omega_2}\sum_{ i_{1'},i_{2'}\atop i_1,i_2}\sum_{\sigma,\sigma'}c^{\dag}_{i_{1'},\sigma}(i\omega_{1'})c^{\dag}_{i_{2'},\sigma'}(i\omega_{2'})V_{\text{charge}}(i\omega_{1'},i_{1'};i\omega_{2'},i_{2'}|i\omega_1,i_1,i\omega_2,i_2)c_{i_2,\sigma'}(i\omega_2)c_{i_1,\sigma}(i\omega_1) \label{srho}
\end{align}
with the general spin-rotationally invariant coupling function
\begin{align}
V_{\text{charge}}(i\omega_{1'},i_{1'};i\omega_{2'},i_{2'}|i\omega_1,i_1,i\omega_2,i_2)=  \sum_{i\nu} f_{\text{charge}}(i\nu ,i_1,i_2) \,   & \delta_{i\omega_1+i\nu,i\omega_{1'}} \delta_{i\omega_2-i\nu,i\omega_{2'}} 
\delta_{i_{1'},i_1}\delta_{i_{2'},i_{2}}  \, . \label{chargepart}
\end{align}
\end{widetext}
Note that the conservation of Matsubara frequencies is implied by the Kronecker-$\delta$s in the last expression.
Hence, for a given pair of sites $(i_1,i_2)$, a density-density interaction with frequency transfer $i\nu$ of the type (\ref{vrhorho})  causes a fixed value $f_{\text{charge}}(i\nu ,i,j)$ of the coupling function  $V_{\text{charge}}(i\omega_{1'},i_{1'};i\omega_{2'},i_{2'}|i\omega_1,i_1,i\omega_2,i_2)$ for all external quantum numbers where the first frequency transfer Mandelstam variable, $i\omega_1- i \omega_{1'}$, is the same. This transfer is the one between those incoming and outgoing frequencies of the legs that need to have the same spin index.  Reversely, if we have a general effective coupling function $V(i\omega_{1'},i_{1'};i\omega_{2'},i_{2'}|i\omega_1,i_1,i\omega_2,i_2)$, e.g. obtained by integrating out some bands, we can average over the two other Mandelstam frequencies $i\omega_1- i \omega_{2'}$ and $i\omega_1+ i \omega_{2}$, or keep these fixed at specific values, in order to obtain an interaction that we can interpret as being of density-density type. 
Looking at the decomposition of the effective interaction in Eq. \ref{Vdecomp}, we could straightaway interpret $\Phi^{\Lambda=1}_{\text{dph}} (i\nu ;\vec{k}_{1'},\vec{k}_{2'},\vec{k}_1;b_{1'},b_{2'},b_{1},b_{2}) $ as density-density contribution, but we will see that a part of this is also needed for the spin-spin part. Later, in a subsequent simplification, we will chose to ignore the  $i\nu$-dependence of that term as well, for being able to use it directly in an interaction Hamiltonian.

We can proceed in a similar way with the spin part. In imaginary time the Heisenberg spin operator is given by
\begin{align}
\vec{S}_i(\tau) &= \frac{1}{2}\sum_{\sigma,\sigma'} c^{\dag}_{i,\sigma}\vec{\sigma}_{\sigma,\sigma'}c_{i,\sigma'}&
\end{align}
\begin{align}
\nonumber
\vec{S}_i(i\nu) &= \int_0^{\beta}d\tau \exp(i\nu\tau)\vec{S}_i(\tau)&\\ 
&= \frac{1}{2\beta}\sum_{\sigma,\sigma'}\sum_{i\omega}c^{\dag}_{i\sigma}(i\omega)\vec{\sigma}_{\sigma,\sigma'}c_{i\sigma'}(i\omega+i\nu)&
\end{align}

A retarded spin-spin interaction term between site-centered spins localized at $i$ and $j$ (again each pair counted once) can be written as
\begin{widetext}
\begin{align}\label{eqspinspininteraction}
S_{\text{spin}} &= \frac{1}{\beta}Ê\sum_{i\nu, (i,j) }Êf_{\text{spin}} (i\nu , i,j) \vec{S}_i(i\nu)  \cdot \vec{S}_j(-i\nu) 
\\Ê
\nonumber
&= -\frac{1}{\beta^3}Ê\sum_{i\nu,  i,j, i\omega , i\omega' \atop \sigma, \sigma'  } f_{\text{spin}} (i\nu ,i,j )  \big[ \frac{1}{2} 
c^{\dag}_{i\sigma}(i\omega)c^{\dag}_{j \sigma'}(i\omega') c_{i\sigma'}(i\omega+ i\nu) c_{j\sigma}(i\omega'-i\nu) \\
&\qquad\qquad\qquad\qquad\qquad\qquad\quad + \frac{1}{4} c^{\dag}_{i\sigma}(i\omega)c^{\dag}_{j \sigma'}(i\omega') c_{j\sigma'}(i\omega' - i\nu) c_{i\sigma}(i\omega +i\nu) \big] \\
& =\frac{1}{\beta^3}\sum_{i\omega_{1'},i\omega_{2'}\atop i\omega_1,i\omega_2}
 \sum_{i_{1'},i_{2'},i_1,i_2 \atop \sigma,\sigma'} 
 c^{\dag}_{i_{1'},\sigma}(i\omega_{1'})c^{\dag}_{i_{2'},\sigma'}(i\omega_{2'})
 V_{\text{spin}}(i\omega_{1'},i_{1'};i\omega_{2'},i_{2'}|i\omega_1,i_1,i\omega_2,i_2)
 c_{i_2,\sigma'}(i\omega_2)c_{i_1,\sigma}(i\omega_1) \, 
\end{align}
now with the coupling function
\begin{align}
\nonumber
V_{\text{spin}}(i\omega_{1'},i_{1'};i\omega_{2'},i_{2'}|i\omega_1,i_1,i\omega_2,i_2)= -
\sum_{i\nu }  f_{\text{spin}} (i\nu,i_{1},i_{2})  & \left[ \frac{1}{2} \,  \delta_{i\omega_1+i\nu,i\omega_{2'}}\delta_{i\omega_2-i\nu,i\omega_{1'}}  \delta_{i_{1'},i_2}\delta_{i_{2'},i_1} \right.
\\
&\left.+\frac{1}{4}\,  \delta_{i\omega_1+i\nu,i\omega_{1'}}\delta_{i\omega_2-i\nu,i\omega_{2'}} \delta_{i_{1},i_{1'}}\delta_{i_{2},i_{2'}} \right.  \label{spinpart}
\end{align}
\end{widetext}
Therefore, the spin part involves a dependence on both Mandelstam frequency transfers. For fixed $i\nu$ we can obtain a relation between the terms in the decomposition (\ref{Vdecomp}) and $f_{\text{charge/spin}} (i\nu ,i,j )$ as follows,
\begin{align}
\Phi^{\Lambda=1}_{\text{dph}}(i\nu;i_1,i_2,i_1,i_2) & = f_{\text{charge}} (i\nu,i_1,i_2) - \frac{1}{4}  f_{\text{spin}} (i\nu,i_1,i_2) &  \\
\Phi^{\Lambda=1}_{\text{crph}}(i\nu;i_2,i_1,i_1,i_2) & = - \frac{1}{2}  f_{\text{spin}} (i\nu,i_1,i_2)  &  \, .
\end{align}
This allows us to understand the data from the fRG from integrating out the high energy bands, given by the functions $\Phi^{\Lambda=1}_{\text{dph}}(i\nu;i_1,i_2,i_1,i_2) $ and $\Phi^{\Lambda=1}_{\text{crph}}(i\nu;i_2,i_1,i_1,i_2) $ as density-density- and spin-spin interacitons $f_{\text{charge/spin}} (i\nu, i,j )$.
These functions have to be added to the bare direct interaction $V_{\mathrm{target-bands}} $ between the low-energy degrees of freedom.

The interpretation becomes simplified if one throws out the frequency dependence of the effective interaction functions $f_{\text{charge/spin}} (i\nu, i,j )$. Then one can formulate an effective interaction Hamiltonian. 
For comparison, let us consider an interaction Hamiltonian of the form
\begin{align}\label{eqgenergalhamiltonian}
\hat{H} = \sum_{(i,j)} U_{ij} \rho_{i}\rho_{j} + \sum_{(i,j)} J_{ij} \vec{S}_i \cdot \vec{S}_j
\end{align}
with instantaneous density-density and spin-spin interactions.  
The corresponding coupling function is, after inserting Grassmann fields for the equal-time density-and spin-operators, given by
\begin{widetext}
\begin{align}\label{eqfrequencystructurevertex}
\nonumber
V(i\omega_{1'},i_{1'};i\omega_{2'},i_{2'}|i\omega_1,i_1,i\omega_2,i_2)=& \sum_{i\nu} \left[Ê\left( U_{i_{1'},i_{2'}}-\frac{1}{4}J_{i_{1'},i_{2'}}\right) \,  \delta_{i\omega_1+i\nu,i\omega_{1'}}\delta_{i\omega_2-i\nu,i\omega_{2'}} \delta_{i_1,i_{1'}}\delta_{i_2,i_{2'}} \right. &\\
&\left. \quad -\frac{1}{2} J_{i_{1'},i_{2'}} \, \delta_{i\omega_1+i\nu,i\omega_{2'}} \delta_{i\omega_2-i\nu,i\omega_{1'}} \delta_{i_1,i_{2'}}\delta_{i_2,i_{1'}} \right] & \, .
\end{align}
\end{widetext}
We can see that this has the same structure as the sum of Eqs. \ref{chargepart} and \ref{spinpart}.
One way to proceed is now to identify \begin{eqnarray}
U_{ij}  & = &  f_{\text{charge}} (i\nu =0,i,j)  \, , \quad \mbox{and}\\
J_{ij} & = & f_{\text{spin}} (i\nu =0,i,j) \, . 
\end{eqnarray}
This will most likely overestimate the effects of the small $i\nu$-contributions, but will still allow to obtain qualitative answers about the tendencies caused by the renormalizations that affect $f_{\text{charge/spin}} (i\nu,i,j)$.
Alternatively one could use $i\nu$-averages of the functions $f_{\text{charge/spin}} (i\nu, i,j )$. 

We note that so far we have restricted the study to local charge and bilinear spin. It is however straightforward to generalize the considerations and identifications above to nonlinear charge and spin fields that live, e.g., on sites $i_1,i_{1'}$. Then we obtain the relations 
\begin{align}
\nonumber
\Phi^{\Lambda=1}_{\text{dph}}(i\nu;i_{1'},i_{2'},i_1,i_{2}) & =  f_{\text{charge}} (i\nu,i_{1'},i_{2'},i_1,i_{2})& \\
\label{eqinterpretationdph}
 &- \frac{1}{4}  f_{\text{spin}} (i\nu,i_{1'},i_{2'},i_1,i_{2})&  \\
\label{eqinterpretationcrph}
\Phi^{\Lambda=1}_{\text{crph}}(i\nu;i_{1'},i_{2'},i_1,i_2) & =  - \frac{1}{2}  f_{\text{spin}} (i\nu,i_{1'},i_{2'},i_1,i_2)   & \, .
\end{align}
The identifications are still taken with respect to the frequency structures.

Finally we can also consider pair-pair interaction of the type
\begin{widetext}
\begin{align}
S_{\text{pairing}} &= \frac{1}{\beta^3}Ê\sum_{i\nu,  i_{1'},i_{2'}, i_1,i_2}\sum_{i\omega , i\omega' \atop \sigma, \sigma'} f_{\text{pairing}} (i\nu ,i_{1'},i_{2'},i_1,i_2 ) 
c^{\dag}_{i_{1'} \sigma}(i\omega')c^{\dag}_{i_{2'} \sigma'}(-i\omega' + i\nu ) c_{i_2\sigma'}(- i\omega+i\nu) c_{i_1\sigma}(i\omega) 
 \\
&=\frac{1}{\beta^3}\sum_{i\omega_{1'},i\omega_{2'}\atop i\omega_1,i\omega_2}\sum_{ i_{1'},i_{2'}\atop i_1,i_2}\sum_{\sigma,\sigma'} c^{\dag}_{i_{1'},\sigma}(i\omega_{1'})c^{\dag}_{i_{2'},\sigma'}(i\omega_{2'})V_{\text{pairing}}(i\omega_{1'},i_{1'};i\omega_{2'},i_{2'}|i\omega_1,i_1,i\omega_2,i_2)c_{i_2,\sigma'}(i\omega_2)c_{i_1,\sigma}(i\omega_1) \label{sP}
\end{align}
with the general spin-rotationally invariant pairing interaction
\begin{align}
V_{\text{pairing}}(i\omega_{1'},i_{1'};i\omega_{2'},i_{2'}|i\omega_1,i_1,i\omega_2,i_2)=  \sum_{i\nu} f_{\text{pairing}}(i\nu ,i_{1'},i_{2'},i_1,i_{2}) \,   & \delta_{-i\omega_1+i\nu,i\omega_{2}} \delta_{-i\omega_{1'}+i\nu,i\omega_{2'}} 
  \, . \label{pairingpart}
\end{align}
\end{widetext}
In the employed vertex decomposition we get
\begin{equation}
\label{eqinterpretationpp}
\Phi^{\Lambda=1}_{\text{pp}}(i\nu;i_{1'},i_{2'},i_1,i_{2})  =   f_{\text{pairing}} (i\nu,i_{1'},i_{2'},i_1,i_2) \, .
\end{equation}
This completes the representation of the effective interaction in terms of charge-charge, spin-spin and pairing interactions.

\section{Results on model systems}\label{results}
Here we describe the numerical results obtained by applying the cfRG scheme described so far to two different few-band models. In the first model, we find only minor deviations from the cRPA results, while in the second example, strong changes occur.
\subsection{Graphene model}

\subsubsection{Hamiltonian}
\begin{figure}[htbp]
    \centering
    \includegraphics[width=0.45\textwidth]{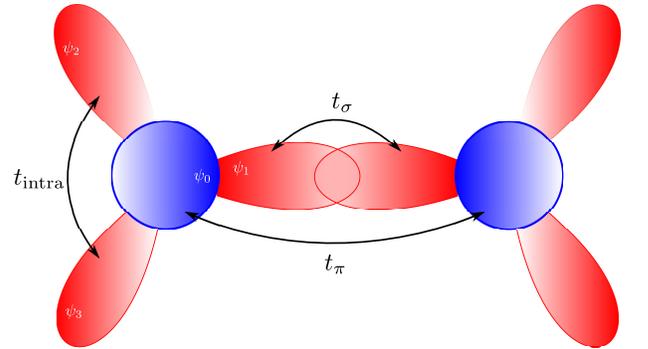}
    \caption[grapheneorbitale]{(color online) $sp^2$ hybridized orbitals in graphene on neighboring C sites. We also indicate the hopping parameters included in the Hamiltonian.} 
    \label{pictorbitale}
\end{figure}

As a first application we discuss in the following a multi-band model of graphene.\cite{Cas10} The unit-cell of a graphene layer is shown in Fig. \ref{pictorbitale}. It consists of two sites with each 3 $sp^2$-hybridized orbitals, forming the $\sigma$-bonds. In addition there is one $p_z$-orbital per lattice site that undergoes $\pi$-bonding  with the $p_z$-orbitals on neighboring sites.  The $\sigma$-bands shall serve as high-energy bands to be integrated out, while the $\pi$-bands are left for the LEEM. These two target bands also feature the well known massless Dirac spectrum at low energies.

The Hamiltonian of our model consists of a free part $\hat{H}_0$ and an interaction part $\hat{H}_{\text{int}}$
\begin{align}
\hat{H}=\hat{H}_0 + \hat{H}_{\text{int}}
\end{align}
The free part is given by
\begin{widetext}
\begin{align}
\nonumber\label{eqnoninthamiltongraphene}
\hat{H}_0&=\sum_{\vec{k}}\bigg(\epsilon_{\pi}\sum_{a=A,B}c^{\dag}_{\vec{k},a,0}c_{\vec{k},a,0}+\epsilon_{\sigma}
\sum_{a=A,B}\sum_{i\neq 0}c^{\dag}_{\vec{k},a,i}c_{\vec{k},a,i}+t_{\text{intra}}\sum_{a=A,B}\sum_{0\neq i \neq j \neq 0}
\left(c^{\dag}_{\vec{k},a,i}c_{\vec{k},a,j}+h.c.\right)\\
&+t_{\pi}\gamma_{\vec{k}}\left(c^{\dag}_{\vec{k},A,0}c_{\vec{k},B,0}+h.c.\right)
+t_{\sigma}\left[e^{i\vec{k}\vec{\delta}_1}c^{\dag}_{\vec{k},A,1}c_{\vec{k},B,1}+
e^{i\vec{k}\vec{\delta}_2}c^{\dag}_{\vec{k},A,2}c_{\vec{k},B,2}+
e^{i\vec{k}\vec{\delta}_3}c^{\dag}_{\vec{k},A,3}c_{\vec{k},B,3}+h.c.\right]
\bigg)
\end{align}
with
\begin{align}
\vec{\delta}_1&=(1,0),\quad \vec{\delta}_2=\frac{1}{2}(-1,\sqrt{3}),\quad \vec{\delta}_3=\frac{1}{2}(-1,-\sqrt{3})\\
\gamma_{\vec{k}}&=\sum_{i=1,2,3}e^{i\vec{k}\vec{\delta}_i}\\
|\gamma_{\vec{k}}|&=\sqrt{3+2\cos\left(\vec{k}(\vec{\delta}_1-\vec{\delta}_2)\right)+2\cos\left(\vec{k}(\vec{\delta}_1-\vec{\delta}_3)\right)
+2\cos\left(\vec{k}(\vec{\delta}_2-\vec{\delta}_3)\right)}
\end{align}
\end{widetext}

\begin{figure}[htbp]
    \centering
    \includegraphics[width=0.45\textwidth]{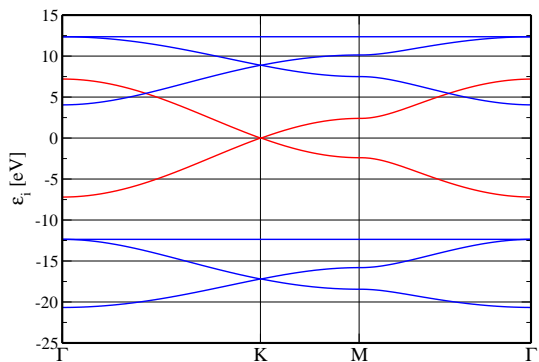}
    \caption[bandstruktur]{(color online) The band-structure of graphene corresponding to the noninteracting Hamiltonian (\ref{eqnoninthamiltongraphene}). There
    are two $\pi$-bands at the Fermi-energy (red) and six high energy $\sigma$-bands (blue).} 
    \label{pictbandstrukturgraphene}
\end{figure}

\begin{figure}[htbp]
    \centering
    \includegraphics[width=0.30\textwidth]{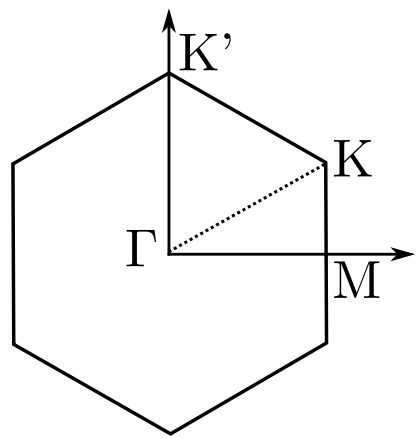}
    \caption[graphenebzone]{The first Brillouin zone of hexagonal lattice for the graphene model.} 
    \label{pictgraphenebzone}
\end{figure}

Here, we choose to compute the values of the parameters above in a Slater-Koster-formalism (see, e.g. Ref. \onlinecite{Mar04}), with localized hydrogen wave-function parametrized by an adjustable Bohr radius, as in Ref.\onlinecite{Cas10}. 
The size of the matrix-elements is given by \cite{Cas10}
\begin{align}
\epsilon_{\pi}&=-11.07eV&\\
\epsilon_{\sigma}&=-13.84eV&\\
t_{\text{intra}}&=-2.77eV&\\
t_{\pi}&=-2.4eV&\\
t_{\sigma}&=-12.36eV&
\end{align}
The resulting band structure with two $\pi$-bands at the Fermi-energy and six high energy $\sigma$-bands is shown in Fig. \ref{pictbandstrukturgraphene}. The $\Gamma$, M and K-points in the first Brillouin-zone are shown in Figure \ref{pictgraphenebzone}.

Note that the matrix elements between $\pi$- and $\sigma$-orbitals are equal to zero. This is due to the mirror symmetry $M_z$ or $z$-reflection symmetry that takes the coordinate perpendicular to the graphene layer, $z$, to $-z$. The $\sigma$ orbitals are even under this symmetry while the $\pi$-orbitals are odd. By a nonzero perpendicular electric field $\vec{E}$ or by deposition of the graphene sheet on a substrate, the $M_z$
symmetry can be broken explicitly. This leads to an additional term in the Hamiltonian of the form
\begin{align}\label{eqhamiltoniangraphenesymmetrybreaking}
\hat{H}_{\pi\sigma}=t_{\pi\sigma}\sum_{\vec{k}}\sum_{a,i\neq0}\left(c^{\dag}_{\vec{k},a,i}c_{\vec{k},a,0}+H.c.\right)
\end{align}

The interaction part is taken as
\begin{align}\label{eqhamiltoniangrapheneinteraction}
\nonumber
\hat{H}_{\text{int}}&=\sum_{n}\sum_{a=A,B}\bigg[U_{\pi\pi}\hat{n}_{n,a,0}\hat{n}_{n,a,0}+U_{\sigma\sigma}\sum_{i\neq0}\hat{n}_{n,a,i}\hat{n}_{n,a,i}&\\
\nonumber
&+U_{\sigma\pi}\sum_{i\neq0}\hat{n}_{n,a,0}\hat{0}_{n,a,i}&\\
&+U_{\text{intra}\sigma}\left(\hat{n}_{n,a,1}\hat{n}_{n,a,2}+\hat{n}_{n,a,1}\hat{n}_{n,a,3}+\hat{n}_{n,a,2}\hat{n}_{n,a,3}\right)\bigg]&
\end{align}
The interaction matrix elements are given by ($V(\vec{r}-\vec{r}')=\frac{1}{|\vec{r}-\vec{r}'|}$, Bohr-radius $a_0=5.29\ 10^{-11}$m).
\begin{widetext}
\begin{align}
\label{upipi}
U_{\pi\pi}&=\int d^3\vec{r}\int d^3\vec{r}' \psi_{0}(\vec{r})^2 V(\vec{r}-\vec{r}') \psi_0(\vec{r}')^2 \approx 0.197\frac{Z^{\ast}}{a_0}&\\
\label{usigsig}
U_{\sigma\sigma}&=\int d^3\vec{r}\int d^3\vec{r}' \psi_i(\vec{r})^2 V(\vec{r}-\vec{r}') \psi_i(\vec{r}')^2 \approx 0.204\frac{Z^{\ast}}{a_0}\quad(i\neq0)&\\
\label{usigpi}
U_{\sigma\pi}&=\int d^3\vec{r}\int d^3\vec{r}' \psi_i(\vec{r})^2 V(\vec{r}-\vec{r}') \psi_0(\vec{r}')^2 \approx 0.171\frac{Z^{\ast}}{a_0}\quad(i\neq0)&\\
\label{uintrasig}
U_{\text{intra}\sigma}&=\int d^3\vec{r}\int d^3\vec{r}' \psi_{1/2/3}(\vec{r})^2 V(\vec{r}-\vec{r}') \psi_{2/3/1}(\vec{r}')^2 \approx 0.156\frac{Z^{\ast}}{a_0}& 
\end{align}
\end{widetext}
with
$\frac{1}{a_0}\mathrel{\hat=}27.2eV$ and $Z^{\ast}=3.25$. $U_{\pi\pi}$ is in good agreement with Ref. \onlinecite{Weh11}.

In principle, the general representation of the Coulomb interaction in the localized-state basis gives rise to many other terms, e.g. such terms with wave function centered at more than two sites. We ignore all other terms that are not of density-density- and two-center-type, as the values for these integrals turn out to be small. Note however that this statement hold only on the 'tree level' of the bare integrals without quantum fluctuations. One of the main points of this paper is that loop-corrections to these parameters can actually generate noticeable terms that are not of density-density type.

\subsubsection{Effective interactions in the $\pi$-sector}

Now our aim is to derive effective interactions between the $\pi$- orbitals by integrating out the high energy $\sigma$-orbitals. This integration can be done
either by using the constrained RPA method or the wick-ordered fRG. The question of our concern is if the latter leads to a qualitatively new picture, i.e. if non-RPA terms are important.

\begin{figure}[htbp]
    \centering
    \includegraphics[width=0.45\textwidth]{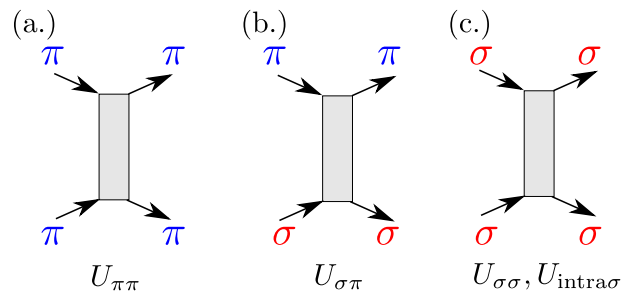}
    \caption[initialcondition]{(color online) The bare one- or two-center interaction vertices taken as nonzero.  The spin projection is conserved along the short side of the rectangle. The interactions can be classified in three types:
    \em{a)} Terms $U_{\pi\pi}$ which couple different $\pi$-orbitals, \em{b)} the terms $\propto U_{\sigma\pi}$ which couple $\sigma$- and $\pi$- orbitals and \em{c)} the terms $\propto U_{\sigma\sigma}, U_{\text{intra}\sigma}$ which
    couple only $\sigma$-orbitals.} 
    \label{pictinitialcondition}
\end{figure}

As argued above the $\pi$- and the $\sigma$-orbitals are not coupled in the free part $\hat{H}_0$ as long as the mirror symmetry $M_z$ holds.  This symmetry decoupling has profound consequences on the perturbation series that is summed in integrating out the $\sigma$-bands.
In Fig. \ref{pictinitialcondition}
we show the bare interaction terms (\ref{upipi}-\ref{uintrasig}), which can be classified in three types: (a.) the terms $\propto U_{\pi\pi}$ which couple $\pi$-orbitals, (b.)
the terms $\propto U_{\sigma\pi}$ which couple $\sigma$- and $\pi$- orbitals and (c.) the terms $\propto U_{\sigma\sigma}, U_{\text{intra}\sigma}$ which
couple only $\sigma$-orbitals.

\begin{figure}[htbp]
    \centering
    \includegraphics[width=0.45\textwidth]{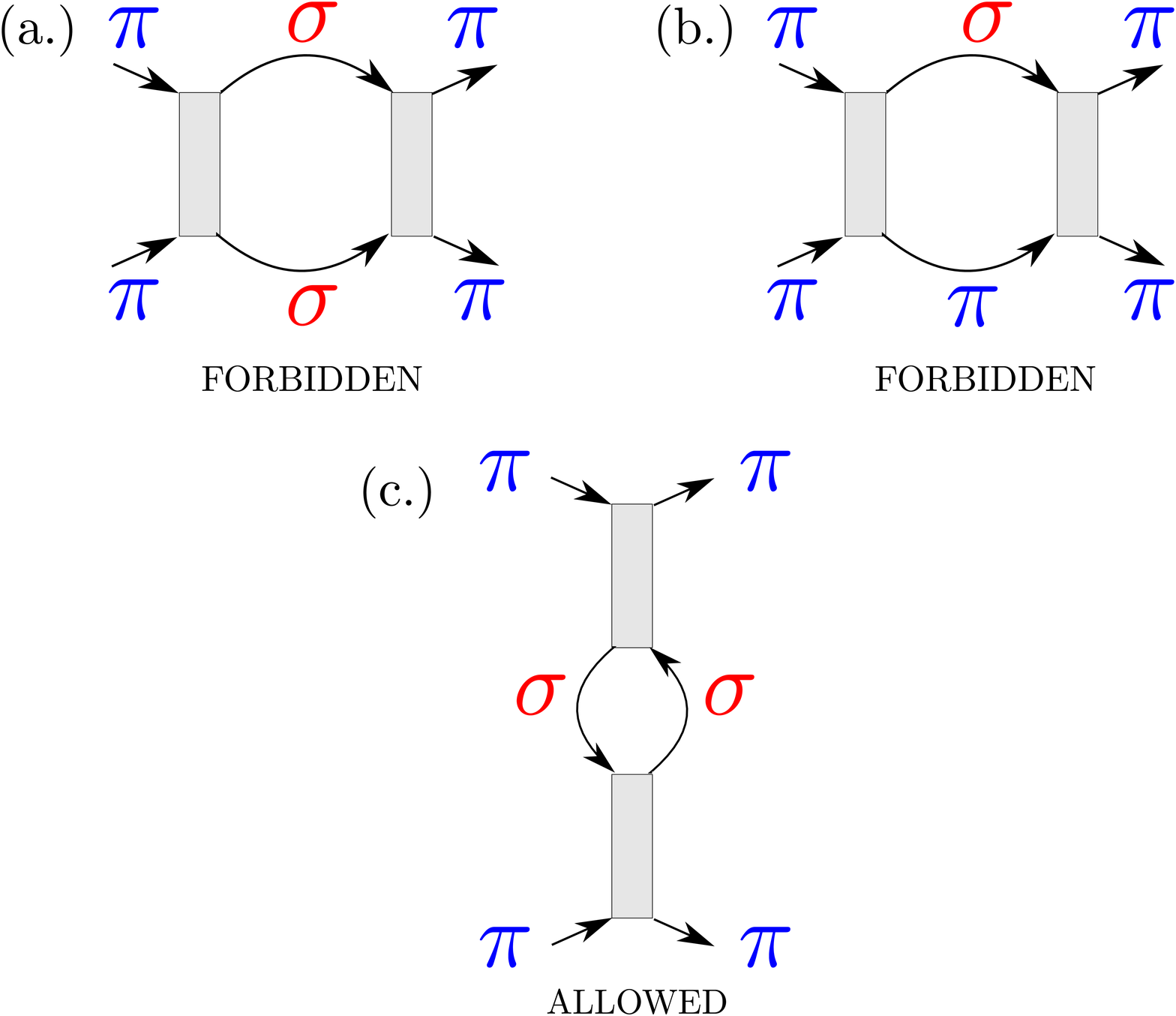}
    \caption[forbiddenterms]{(color online) Different diagrams of second order as appearing on the right hand side of the fRG equation for the effective interactions. Due to the symmetry-restricted set of initial interactions, the only nonzero contribution with 4 external $\pi$-band legs comes form the RPA-type diagram in (c). The contributions from the particle-particle diagram, from the crossed particle hole diagram and from the vertex correction diagrams in the direct particle hole channel cancel.  } 
    \label{pictforbiddenterms}
\end{figure}

Now we consider the effective interactions in the $\pi$-sector in second order perturbation theory, with the internal lines constrained to contain at least one propagator in $\sigma$-sector. In Fig.~\ref{pictforbiddenterms} \em{ a)} and \em{b)}
we show possible contributions to the effective interaction from the particle-particle diagram. As the $\sigma$-orbitals are integrated out at least one inner line of this
diagram has to belong to the $\sigma$-sector. But as shown in Fig.~\ref{pictinitialcondition}, the bare interaction vertices corresponding to the 
Hamiltonian (\ref{eqhamiltoniangrapheneinteraction}) have a different structure than the allowed interaction vertices in Fig.~\ref{pictforbiddenterms}.
Therefore we do not get any nonzero  contributions to the effective $\pi\pi$-interaction from the particle-particle channel in second order perturbation theory.
A completely analogous argumentation holds for the crossed-particle hole channel and the last two 'vertex correction' diagrams of the direct particle-hole channel (cf. Fig.~\ref{pictflowequations}). Of course the vanishing of this contributions depends on neglecting bare interactions other than of density-density type, but if these were nonzero,  only minor corrections would occur. In principle, processes within the high-energy bands can also these non-density-density terms, as will also be visible in our data. But these terms just lead to small corrections, as their magnitude is controlled by the energy gap $\Delta E$ between the target bands and the high-energy bands.

Thus it turns out that the only significant renormalization that is generated in second order perturbation theory with the bare interaction
vertices from Hamiltonian (\ref{eqhamiltoniangrapheneinteraction}) results from the first direct particle-hole diagram shown in Fig.~\ref{pictforbiddenterms} {\em c)}. 
This diagram is of course also included in the cRPA approach, i.e. on this level of approximation both approaches are equivalent.

Beyond second order perturbation theory, the picture becomes slightly more difficult. For example interactions exclusively in the $\sigma$-sector that can be generated from
all channels can be inserted into the inner part of the diagram of Fig.~\ref{pictforbiddenterms} {\em c}. However, the numerical analysis for the realistic parameters listed above does not reveal sizable deviations from the cRPA. 

\begin{figure*}[htbp]
    \centering
    \includegraphics[width=0.45\textwidth]{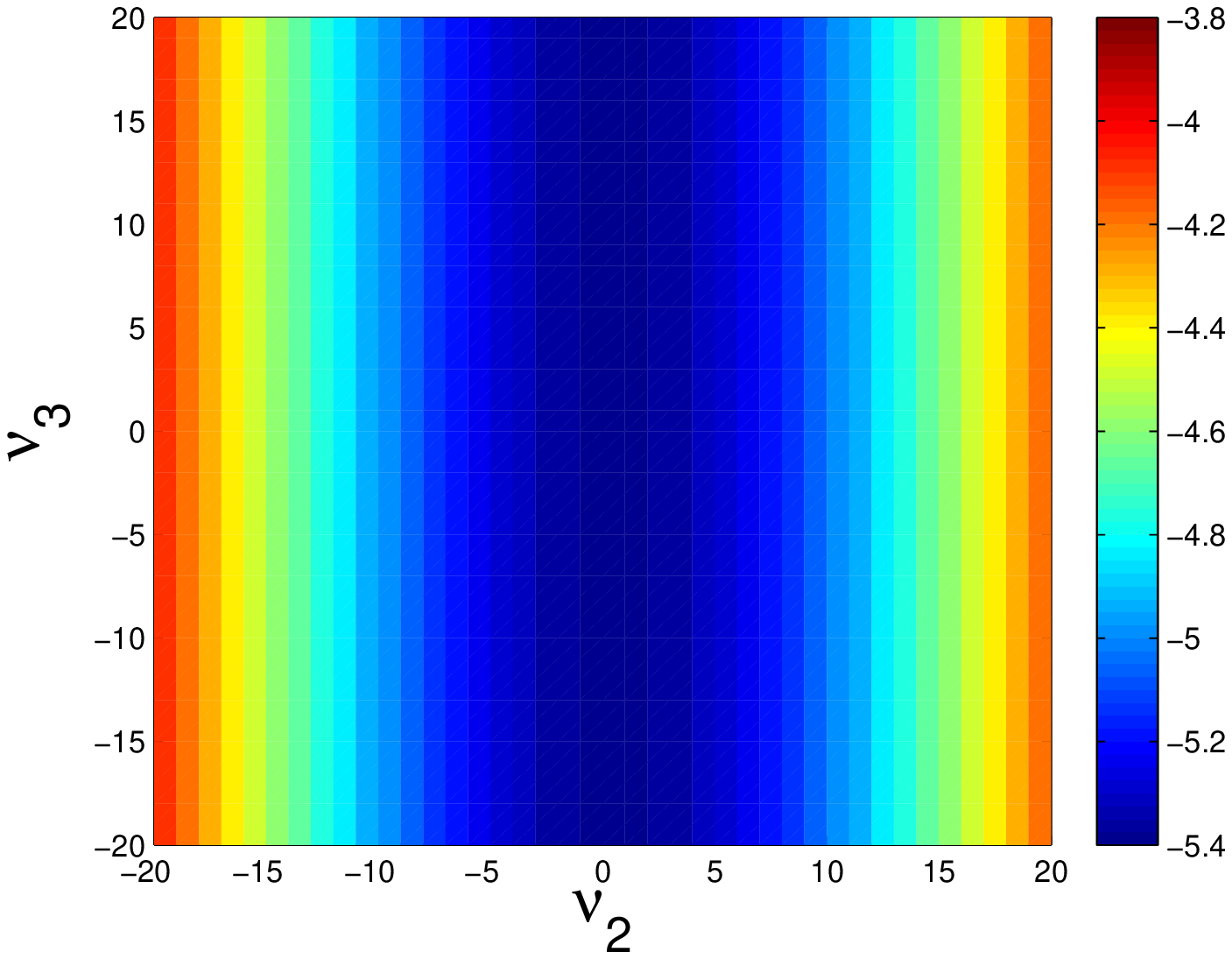}
    \includegraphics[width=0.45\textwidth]{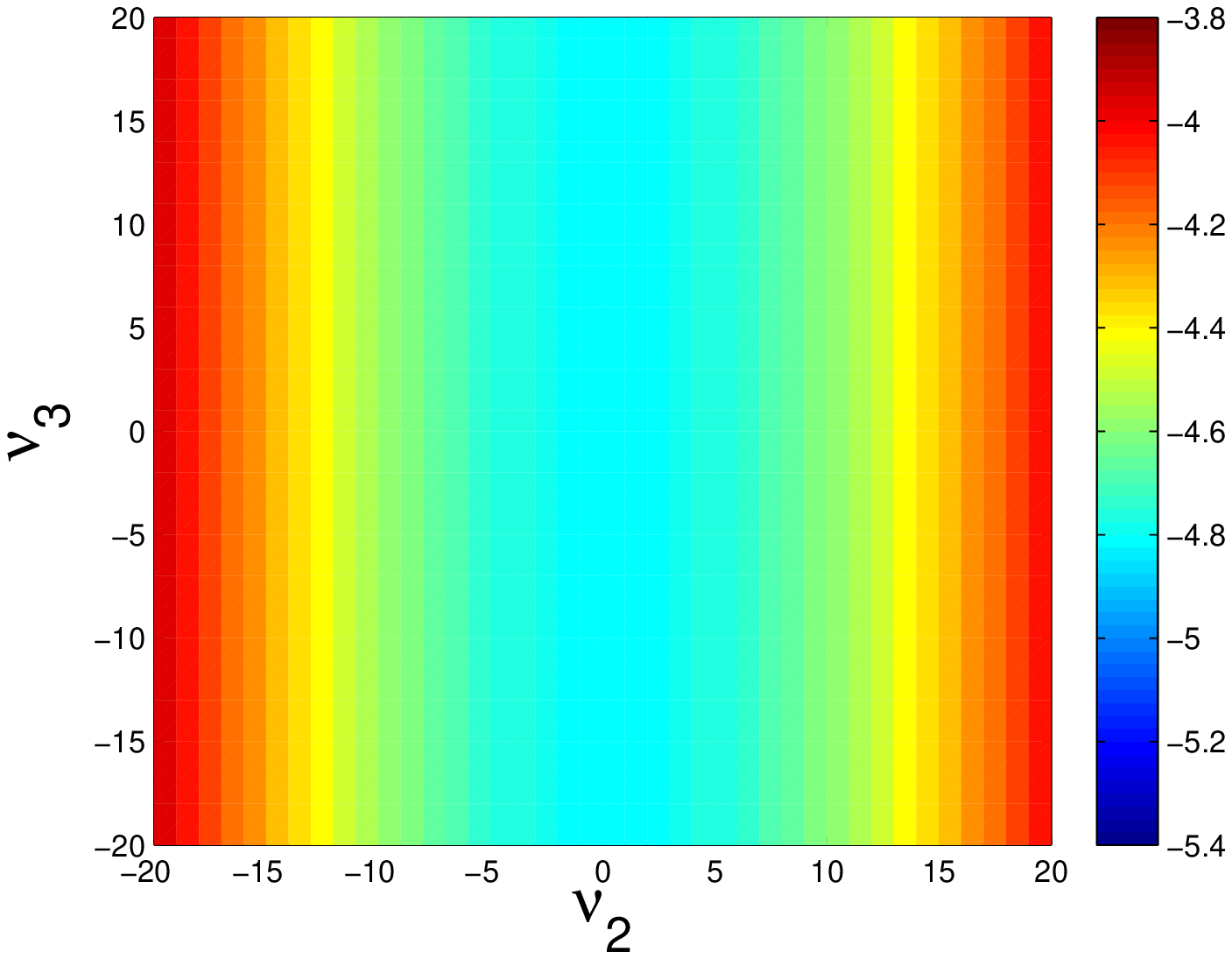}
    \caption[graphenevertices]{(color online) The vertex $V(\vec{K},\pi_{A};\vec{K},\pi_{A}|\vec{K},\pi_{A};\vec{K},\pi_{A})-U$ in cfRG (left) and cRPA (right).} 
    \label{pictgraphenevertices}
\end{figure*}

In Fig.~\ref{pictgraphenevertices} we show data for the effective interactions in the target bands, for the case where all external legs are chosen at one $K$-point in the same $\pi$-band. It can be clearly seen that the differences of the cfRG to the cRPA are at best of minor quantitative type. In the cRPA calculation the vertex only depends on the second transfer frequency $i\nu_2$ corresponding to the frequency dependence generated by the first direct particle-hole bubble. In the cfRG calculation no additional frequency dependence on $i\nu_1$ or $i\nu_3$ is generated and a peaked structure depending on $i\nu_2$ is observed. The structure is qualitatively the same as in the cRPA calculation.

Of course, this is only a statement about a single point in momentum space, but it illustrates the diagrammatic argument that non-cRPA corrections are weak. 
Thus we can conclude that also in more ab-initio based studies for this system like done in Ref. \onlinecite{Weh11}, the result of the cRPA-analysis should come very close to the answer one would obtain when more than the cRPA diagrams are considered.

This picture can be expected to change parametrically when we include a symmetry-breaking term (\ref{eqhamiltoniangraphenesymmetrybreaking}). 
Then, the high-energy bands will obtain admixtures of the $\pi$-orbitals and vice versa. This opens new interaction channels, as now the formerly forbidden terms lead to contributions parametrically large or small in the strength of the symmetry-breaking. This situation will occur e.g. in layered graphene, e.g. in bilayers, or for the outer layers or asymmetric bands of trilayer systems. Similarly, putting monolayers on a substrate or electric field in $z$-direction will violate the $z$-reflection symmetry as well. 

We here try to capture these effects qualitatively by introducing a nonzero hopping $t_{\pi\sigma}$ that now couples the two sets of bands.

\begin{figure}[htbp]
    \centering
    \includegraphics[width=0.50\textwidth]{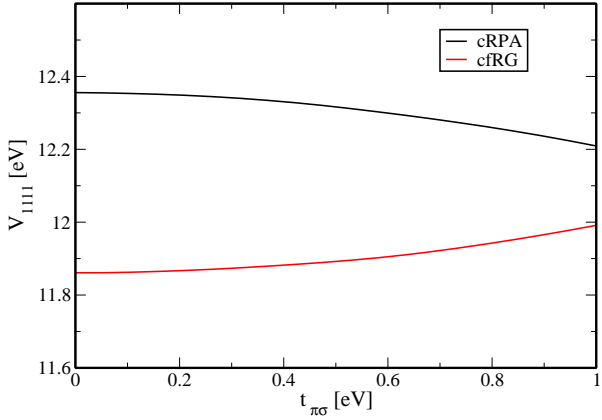}
    \caption[graphenewechselwirkungcomparison]{(color online) Comparison of the fRG and the cRPA calculation of the matrix-element 
    $V_{1111}\equiv V(\vec{K},\pi_{A};\vec{K},\pi_{A}|\vec{K},\pi_{A};\vec{K},\pi_{A})$ at zero frequency transfer as function of the hopping
    $t_{\pi\sigma}$.} 
    \label{pictgraphenewechselwirkungcomparison}
\end{figure}
\begin{figure}[htbp]
    \centering
    \includegraphics[width=0.50\textwidth]{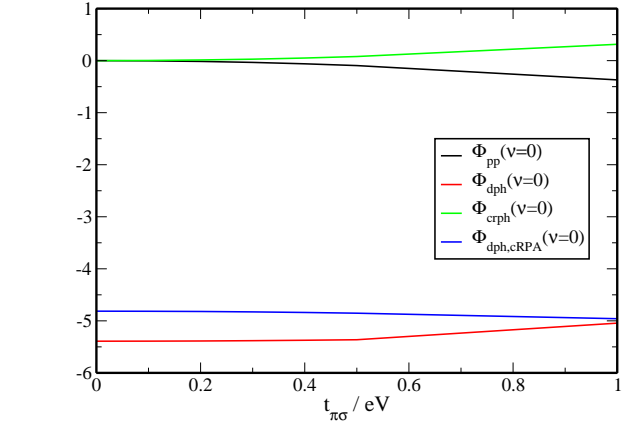}
    \caption[channel1111]{(color online) Comparison of the three interaction channels $\Phi_{i}\equiv V(\vec{K},\pi_{A};\vec{K},\pi_{A}|\vec{K},\pi_{A};\vec{K},\pi_{A})$ 
      at zero frequency transfer for $i=\text{pp},\text{dph},\text{crph}$ with the cRPA calculation.} 
    \label{pictchannel1111}
\end{figure}

In Fig.~\ref{pictgraphenewechselwirkungcomparison} we compare the fRG and the cRPA calculation of the coupling function component $V_{1111}\equiv V(\vec{K},\pi_{A};\vec{K},\pi_{A}|\vec{K},\pi_{A};\vec{K},\pi_{A})$ in the target $\pi$-bands at zero frequency transfer as function of 
$t_{\pi\sigma}$. $\vec{K}$ is one of the Dirac-Points shown in Fig. \ref{pictgraphenebzone}.
The bare matrix element $V_{1111}^{\Lambda=0}=17.4 eV$ becomes screened both in the cRPA and in the fRG calculation. For $t_{\pi\sigma}=0$ the screening effect is larger
in the fRG calculation than in cRPA, which is caused by the corrections beyond second order perturbation theory.  For $t_{\pi\sigma}\neq 0$ a different behavior in both approximations
is observed. In the cRPA calculation, the screening becomes larger when $t_{\pi\sigma}$ is increased, while in the fRG calculation we obtain an anti-screening of the interactions
which is mainly caused by the crossed particle-hole channel (cf. the channel-decomposed screening behavior in Fig. \ref{pictchannel1111} that shows the increase in the crossed particle-hole channel coincident with less reduction in the direct particle-hole channel). Note however that this different behavior with $t_{\pi\sigma}$ becomes significant at values of $t_{\pi\sigma}$ that are
very large compared to the bare energy scales of the Hamiltonian. Furthermore, at $t_{\pi\sigma}\not= 0$, the corrections to cRPA make the cfRG-result more similar to the cRPA data.  So, in the present example, the cRPA results appear to be quite robust even if the two sets of bands are allowed to couple.

An important learning from this study is however that the symmetry of a problem can decide if there are corrections to cRPA or not. If the $z$-reflection symmetry is broken, new terms enter the perturbation series for the effective interaction which contain physics that go beyond screening arguments. In principle this opens a new variant of tuning the interaction parameters of a two-dimensional system, although the precise impact of the symmetry breaking is possibly not intuitively clear.

\subsection{One-dimensional chain}

\subsubsection{Hamiltonian}

\begin{figure}[htbp]
    \centering
    \includegraphics[width=0.45\textwidth]{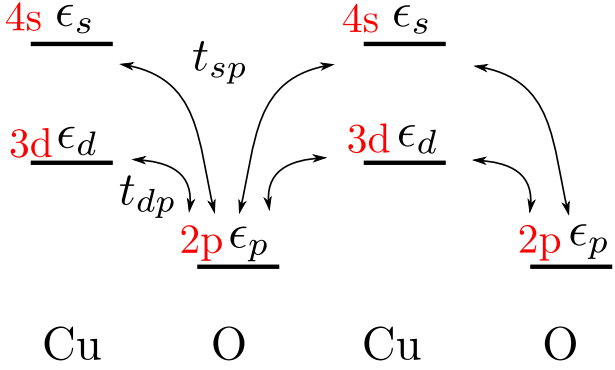}
    \caption[cuochain]{(color online) The one-dimensional chain consists of 'copper' sites with $4s$- and $3d$-orbitals and 'oxygen' sites with $2p$-orbitals. The hopping matrix element between the $s$- and $d$-orbitals on the same site is equal to zero due to symmetry.} 
    \label{pictcuochain}
\end{figure}

As a second example we consider a simplistic model for  one-dimensional chain, inspired by copper oxides. It consists of 'copper'-centered $4s$- and $3d$-orbitals and 'oxygen' sites with $2p$-orbitals extending towards the copper sites (cf. Fig. \ref{pictcuochain}). This model can be considered as more generic than the graphene example as there is no symmetry difference between the bands near and further away form the Fermi level. 

The Hamiltonian is given by
\begin{align}
\label{eqhamiltoncuochain}
\hat{H} &= \hat{H}_0+\hat{H}_{\text{int}}&\\
\nonumber
\hat{H}_0 &= \epsilon_s \sum_{k,\sigma}s^{\dag}_{k,\sigma}s_{k,\sigma}
+ \epsilon_d \sum_{k,\sigma}d^{\dag}_{k,\sigma}d_{k,\sigma}
+ \epsilon_p \sum_{k,\sigma}p^{\dag}_{k,\sigma}p_{k,\sigma}&\\
\nonumber
&+ t_{sp} \sum_{k,\sigma}\left[1+\exp(ik)\right]\left(s^{\dag}_{k,\sigma}p_{k,\sigma}+H.c.\right)&\\
&+ t_{dp} \sum_{k,\sigma}\left[1+\exp(ik)\right]\left(d^{\dag}_{k,\sigma}p_{k,\sigma}+H.c.\right)&
\end{align}
\begin{align}
\nonumber
\hat{H}_{\text{int}} &= U_{s}\sum_{i}s^{\dag}_{i,\uparrow}s_{i,\uparrow}s^{\dag}_{i,\downarrow}s_{i,\downarrow}
+ U_{d}\sum_{i}d^{\dag}_{i,\uparrow}d_{i,\uparrow}d^{\dag}_{i,\downarrow}d_{i,\downarrow} & \\
&+ U_{p}\sum_{i}p^{\dag}_{i,\uparrow}p_{i,\uparrow}p^{\dag}_{i,\downarrow}p_{i,\downarrow}
+ U_{sd} \sum_{i,\sigma,\sigma'}s^{\dag}_{i,\sigma}s_{i,\sigma}d^{\dag}_{i,\sigma'}d_{i,\sigma'} &. \, 
\end{align}
We choose 
$\epsilon_s = 2$, $\epsilon_d = -0.5$, $\epsilon_p=-2$, $t_{sp}=1$, $t_{dp}=1$, $U_s=U_p=U_d=U=1.5$, $U_{sd}=U/3$, all measured the same energy unit. The temperature is chosen as $\beta=1/T=5$.

\begin{figure}[htbp]
    \centering
    \includegraphics[width=0.50\textwidth]{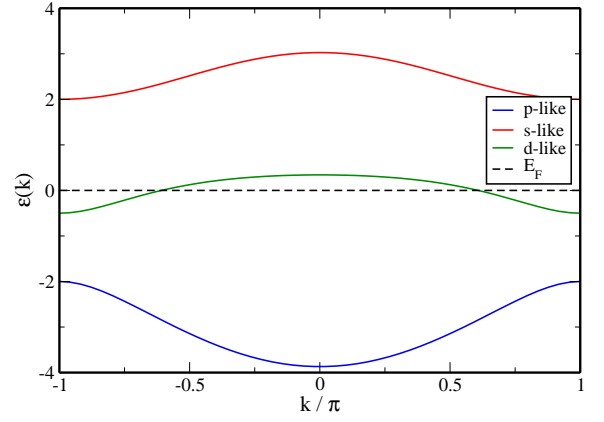}
    \caption[energien1dkette]{(color online) The spectrum of the free part $\hat{H}_0$ of the Hamiltonian (\ref{eqhamiltoncuochain}). It consists of two valence bands and one conduction band. They are called $s$-like, $d$-like and $p$-like according to
    their main orbital contributions.} 
    \label{pictenergien1dkette}
\end{figure}

The hopping element between the copper $4s$- and $3d$-orbital is equal to zero, due to symmetry. The spectrum of the free part $\hat{H}_0$ is
shown in Fig.~\ref{pictenergien1dkette}. It consists of two high-energy bands above and below the Fermi level and one conduction band. They are called $s$-like, $d$-like and $p$-like according to
their main orbital contribution. With parameters listed above, the $d$-like is (roughly) half filled. 

In the following we integrate out the $p$-like and the $s$-like band to get
an effective theory for the $d$-like band. Note that now, due to the overlaps $t_{dp}$ between the $d$-orbitals and the $p$-orbitals, the $d$-orbital also has some weight in the high-energy bands. This opens the way for non-RPA contributions in the integration over the high-energy bands. Hence we expect stronger deviation of the cfRG from the cRPA.

\subsubsection{Effective Interactions}
As the model is one-dimensional, we can also take first steps to explore the wavevector dependence of the effective interactions. We do this by dividing up the one-dimensional Brillouin zone in $N$ patches. Here we use $N=2$ and $N=4$, but with slightly higher numerical ambitions, this number can be increased.  To be more precise,  in a Scheme 1 we use two $k$-points at $k=0$ and $k=\pi$ and 41 Matsubara frequencies.
In a Scheme 2 we use the $k$-points $k=-\pi/2$, $k=0$, $k=\pi/2$ and $k=\pi$ and 11 Matsubara frequencies. We restricted the internal $k$-sums in the flow equations also to this discrete set of $k$-points, which is a valid approximation  due to the relatively weak dispersion of the bands compared to the band gaps. So, technically, our calculation corresponds to a finite size cluster and by Fourier transformation we can obtain local and nonlocal effects up to three neighbored lattice sites.

\begin{figure*}[htbp]
    \centering
    \includegraphics[width=0.4\textwidth]{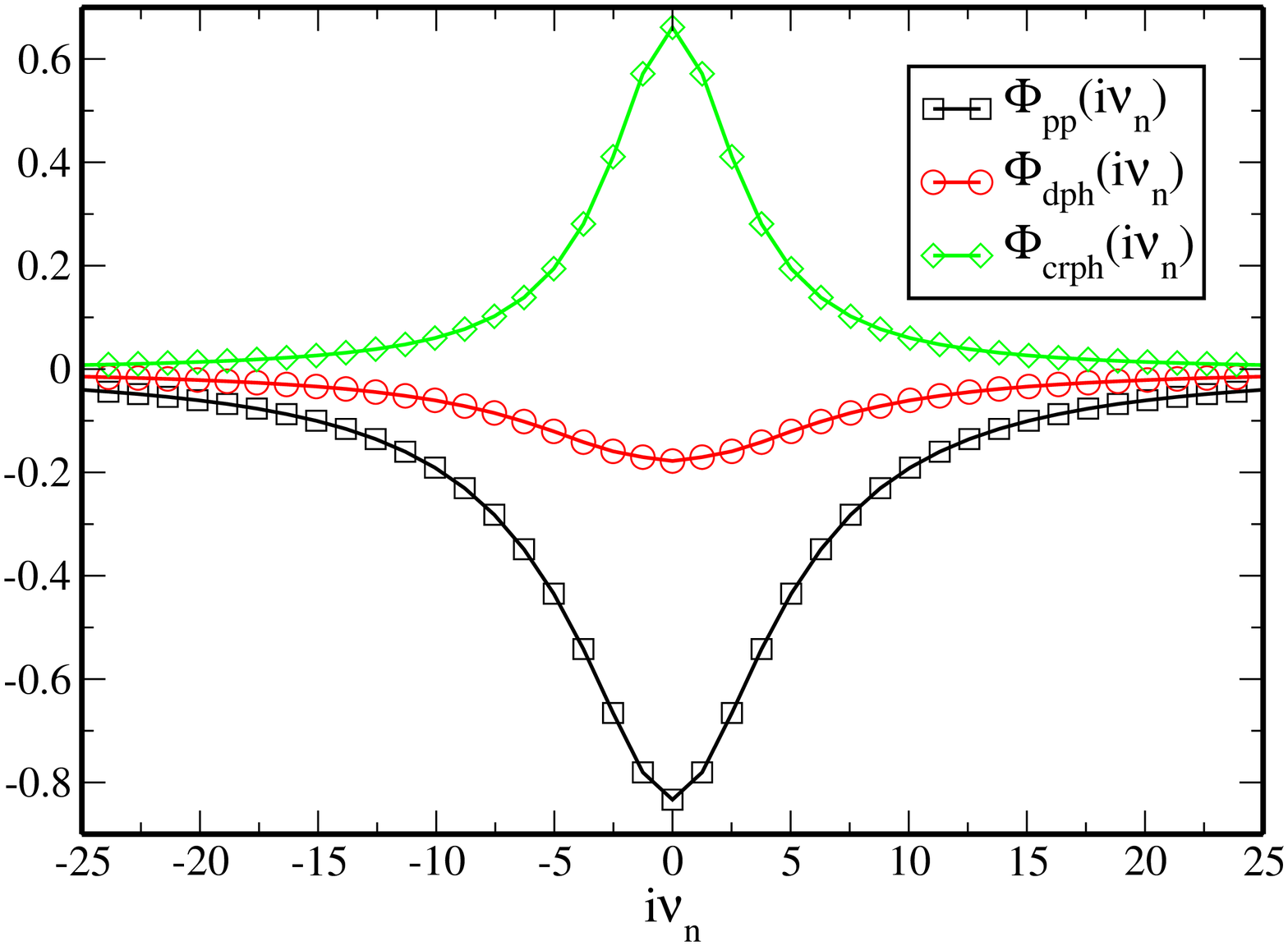}
    \includegraphics[width=0.4\textwidth]{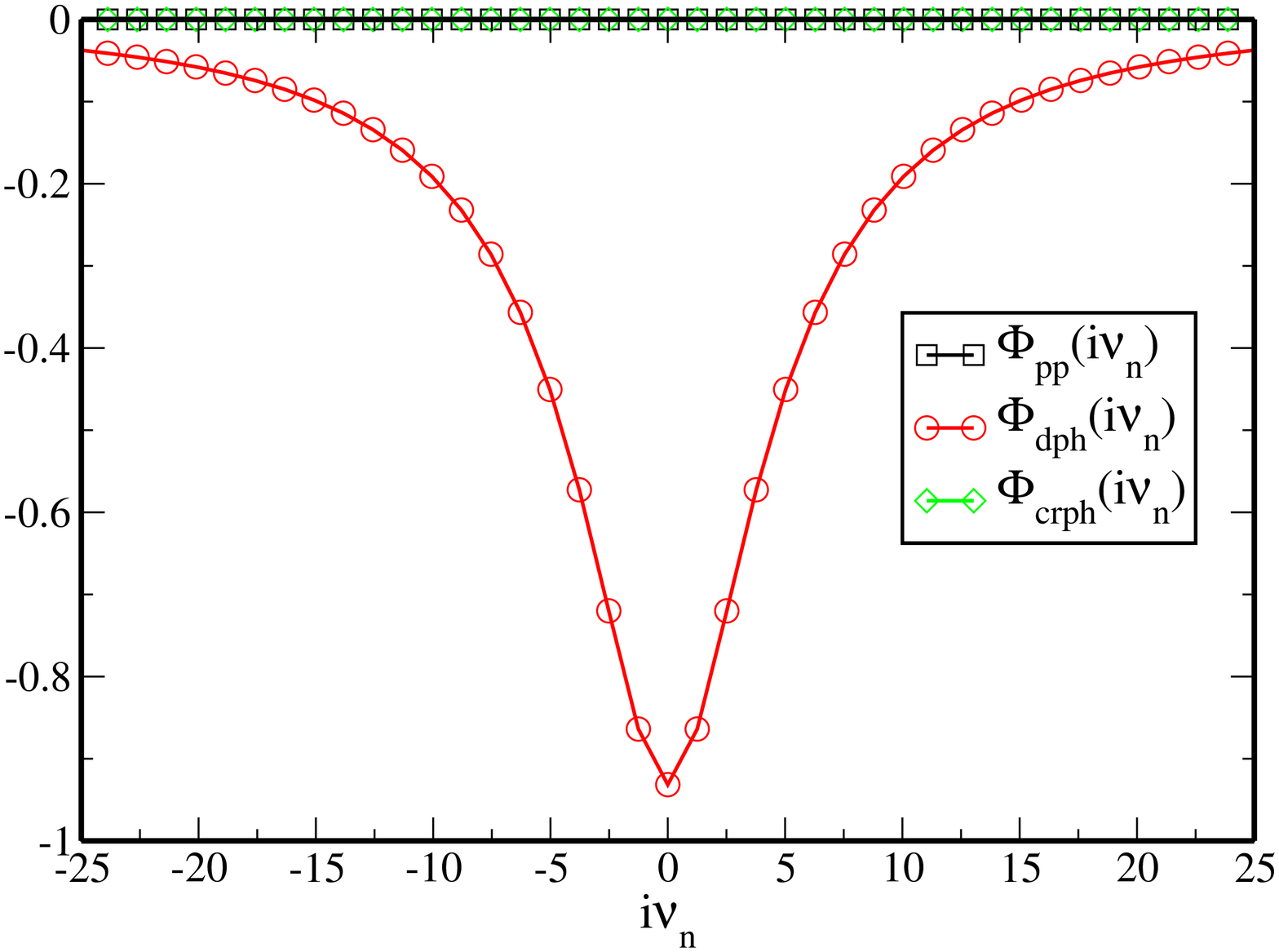}\\
    \includegraphics[width=0.4\textwidth]{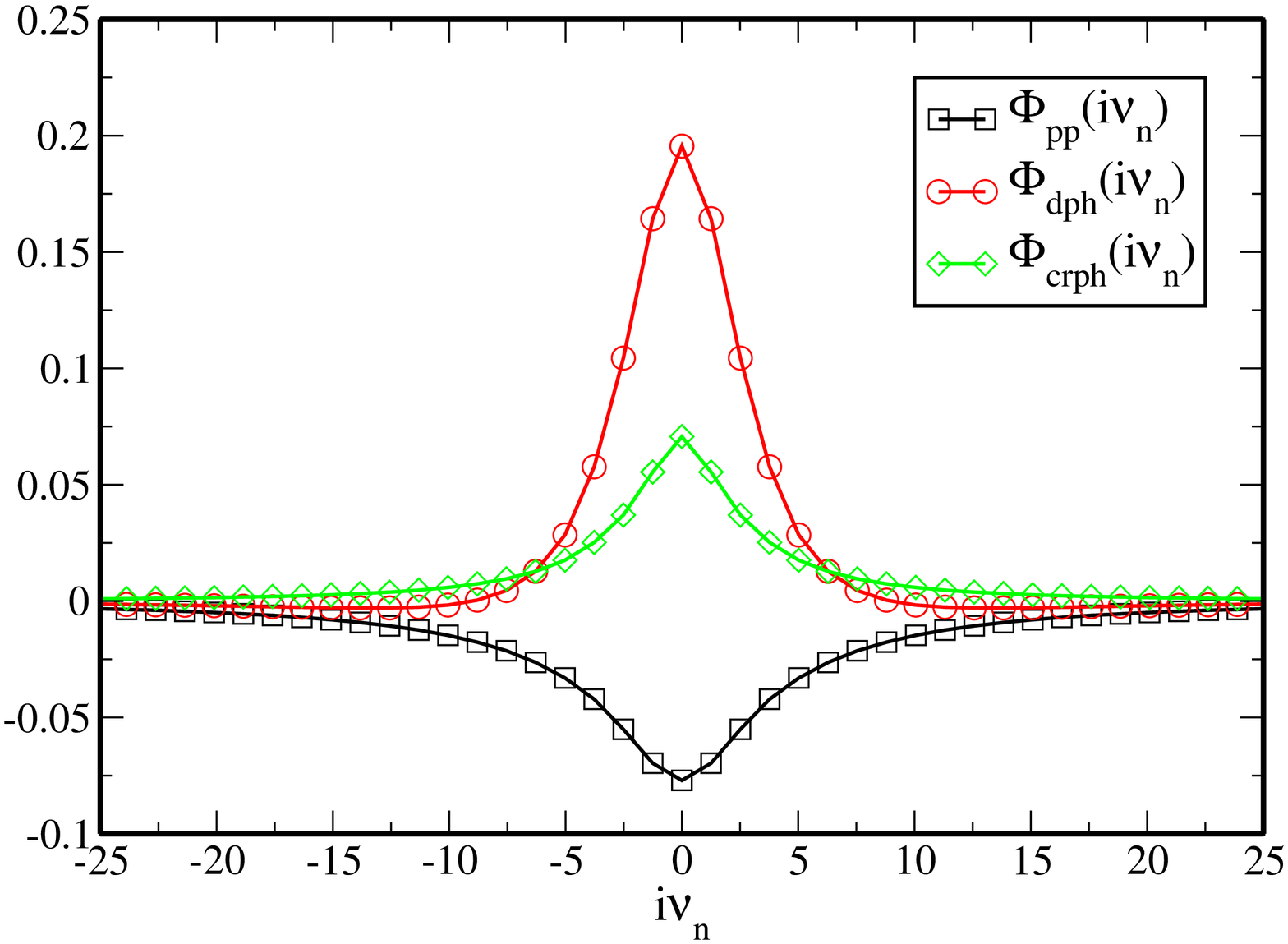}
    \includegraphics[width=0.4\textwidth]{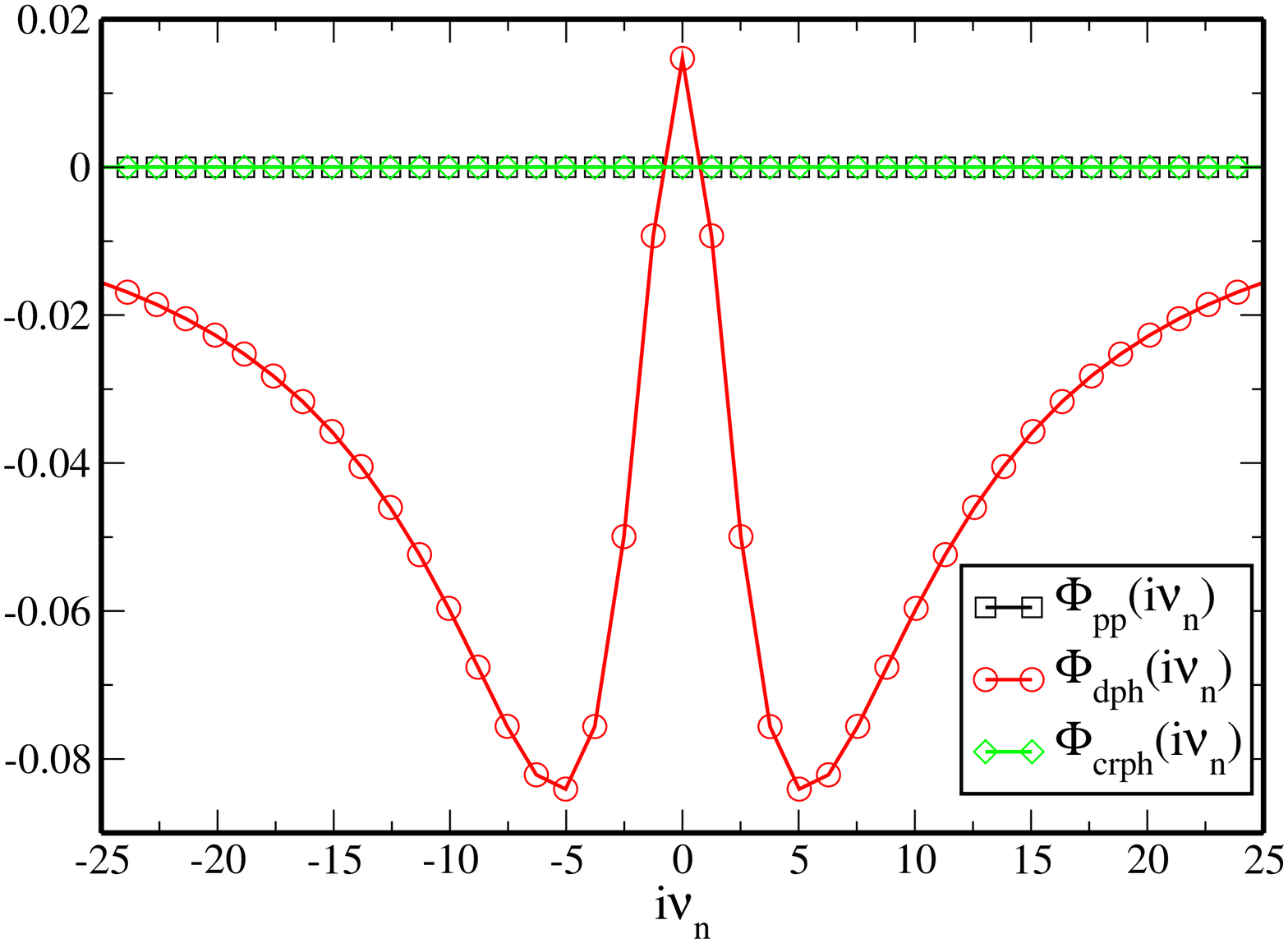}\\
    \includegraphics[width=0.4\textwidth]{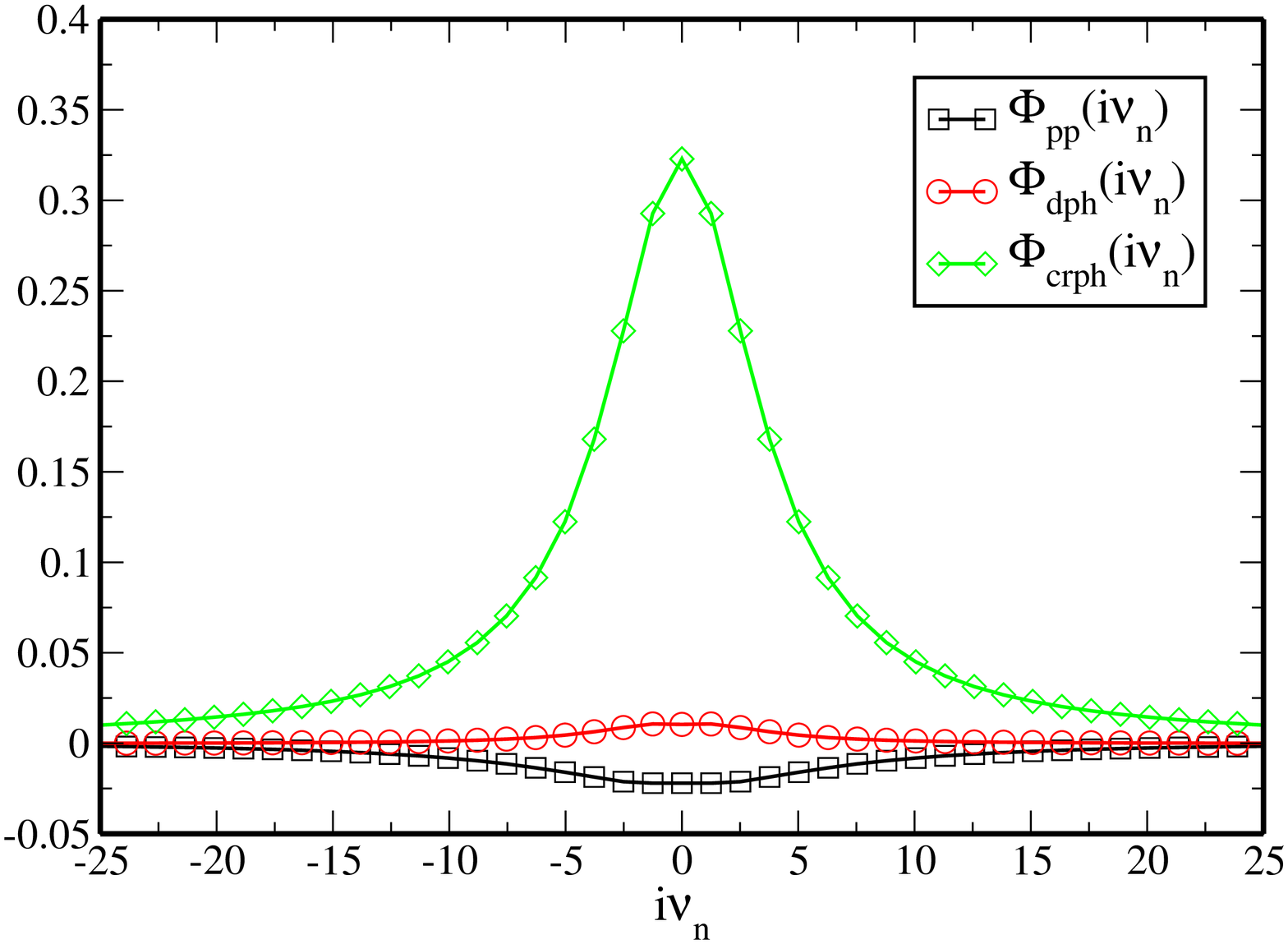}
    \includegraphics[width=0.4\textwidth]{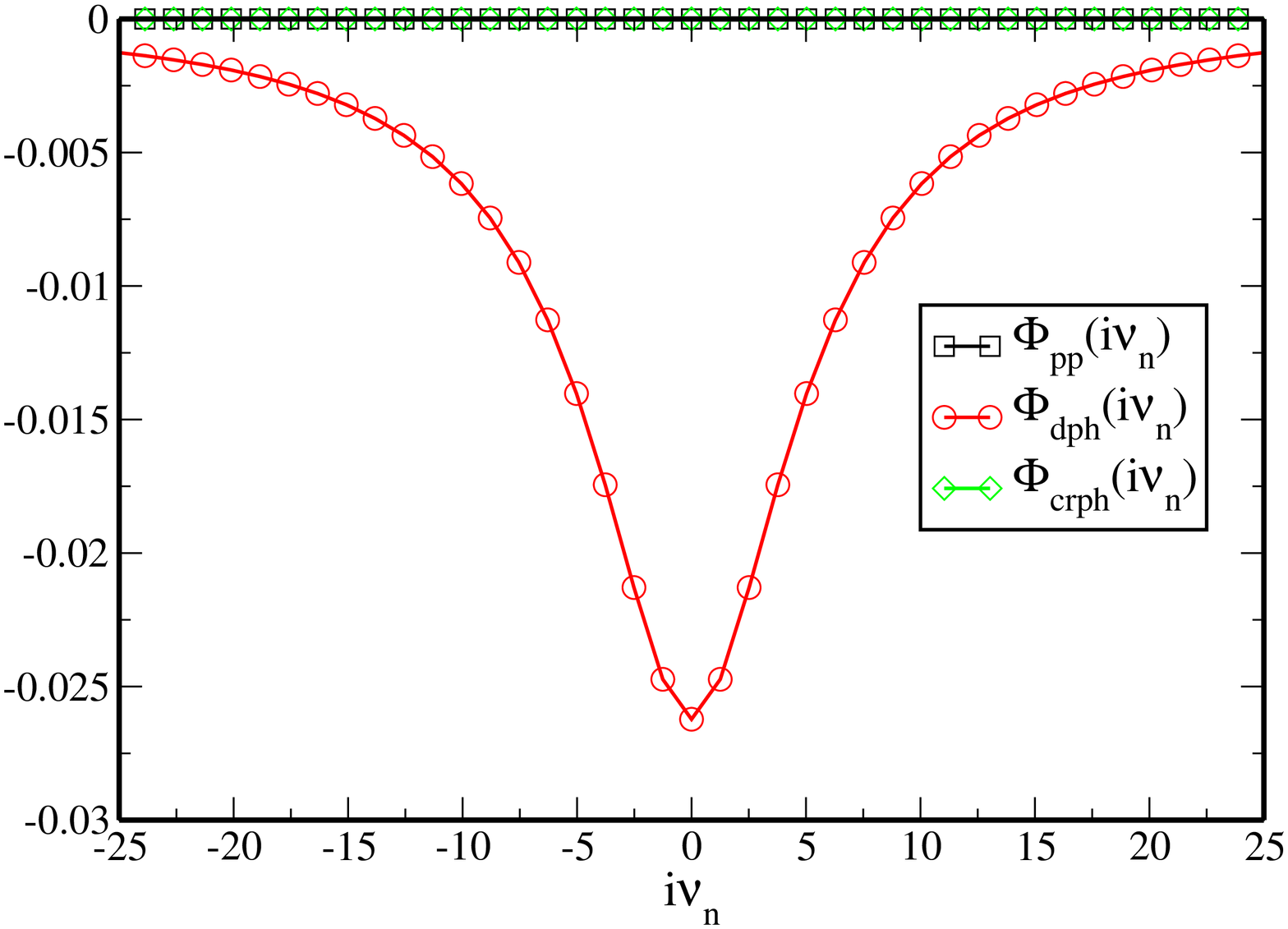}
    \caption[vertex]{(color online) Effective $d$-band interactions at $U=3$, in units of $t_{sp}$. First row: The three interaction channels $\Phi_{i}^{\Lambda=1}(i\nu;1,1,1,1;d,d,d,d)$ with $i=\text{pp},\text{dph},\text{crph}$ in fRG (left) and cRPA (right). The indices '$1,1,1,1$' refer to lattice sites, so this is the local interaction.
    Second row: The three interaction channels $\Phi_{i}^{\Lambda=1}(i\nu;1,2,1,2;d,d,d,d)$ with $i=\text{pp},\text{dph},\text{crph}$ in fRG (left) and cRPA (right). The site combination '$1,2,1,2$' is such that the spin projections on site 1 and 2 remain both unchanged in the interaction process.
    Third row: The three interaction channels $\Phi_{i}^{\Lambda=1}(i\nu;1,2,2,1;d,d,d,d)$ with $i=\text{pp},\text{dph},\text{crph}$ in fRG (left) and cRPA (right). The indices 1 and 2 refer to neighboring lattice sites. The site combination '$1,2,1,2$' is such that the spin projections on site 1 and 2 are exchanged in the interaction process if the incoming spin projection is zero.}
    \label{pictvertex}
\end{figure*}

In Fig.~\ref{pictvertex} we show the three interaction channels $\Phi_{i}^{\Lambda=1}(i\nu;i,j,k,l;d,d,d,d)$ with $i=\text{pp},\text{dph},\text{crph}$ in position space, comparing cfRG in the left plots with cRPA in the right plots. Of course, in the cRPA only the direct particle-hole channel contributes by construction.
Note that $\Phi_{\text{crph}}(i\nu=0)$ is positive in all cases in the cfRG, which corresponds with Eq.(\ref{eqfrequencystructurevertex}) to a negative (i.e. ferromagnetic) exchange interaction $J$.
Below we will show that this generated exchange interaction leads to a polarized ground state at least in a small system. In the cRPA calculation one has $\Phi_{\text{crph}}(i\nu)\equiv 0$ and
therefore the effective $J$ according to our decomposition of the effective interaction is zero. The nearest-neighbor direct particle-hole term in the second row has a distinct behavior as a function of the transfer frequency due to a sign change between the $k=0$ and the $k=\pi$ term. This is already visible
in second order perturbation theory.

\begin{figure*}[htbp]
    \centering
    \includegraphics[width=0.4\textwidth]{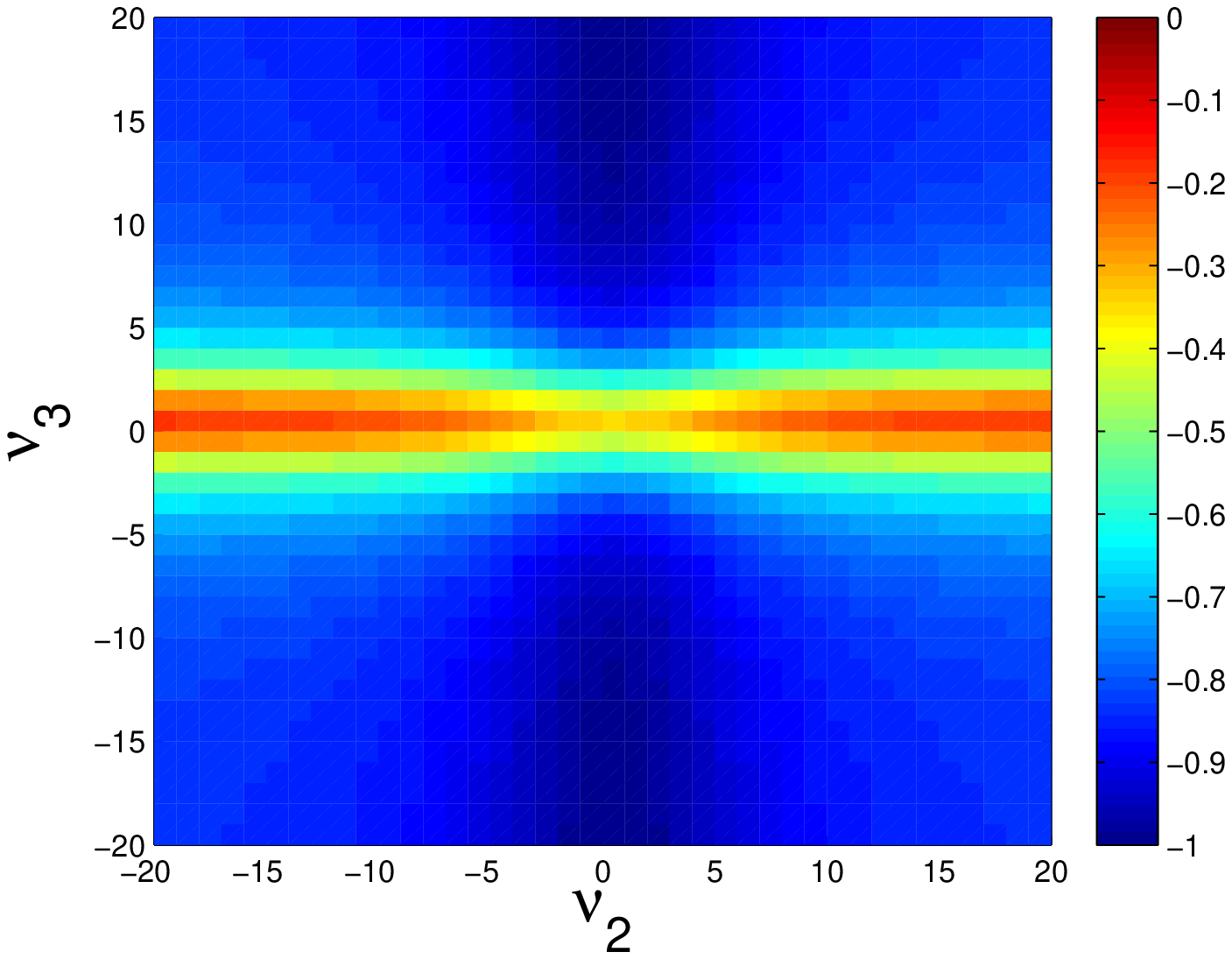}
    \includegraphics[width=0.4\textwidth]{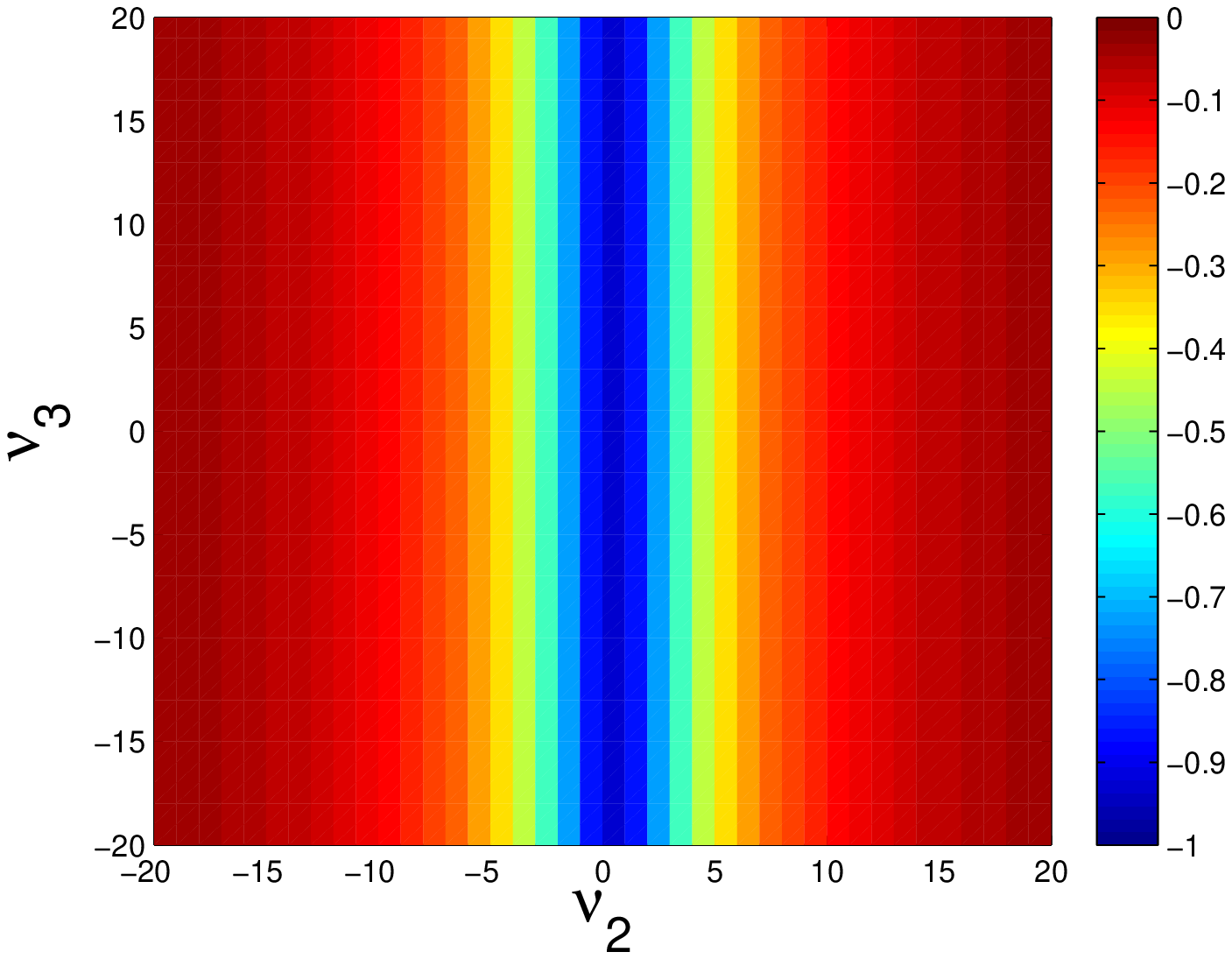}\\
    \includegraphics[width=0.4\textwidth]{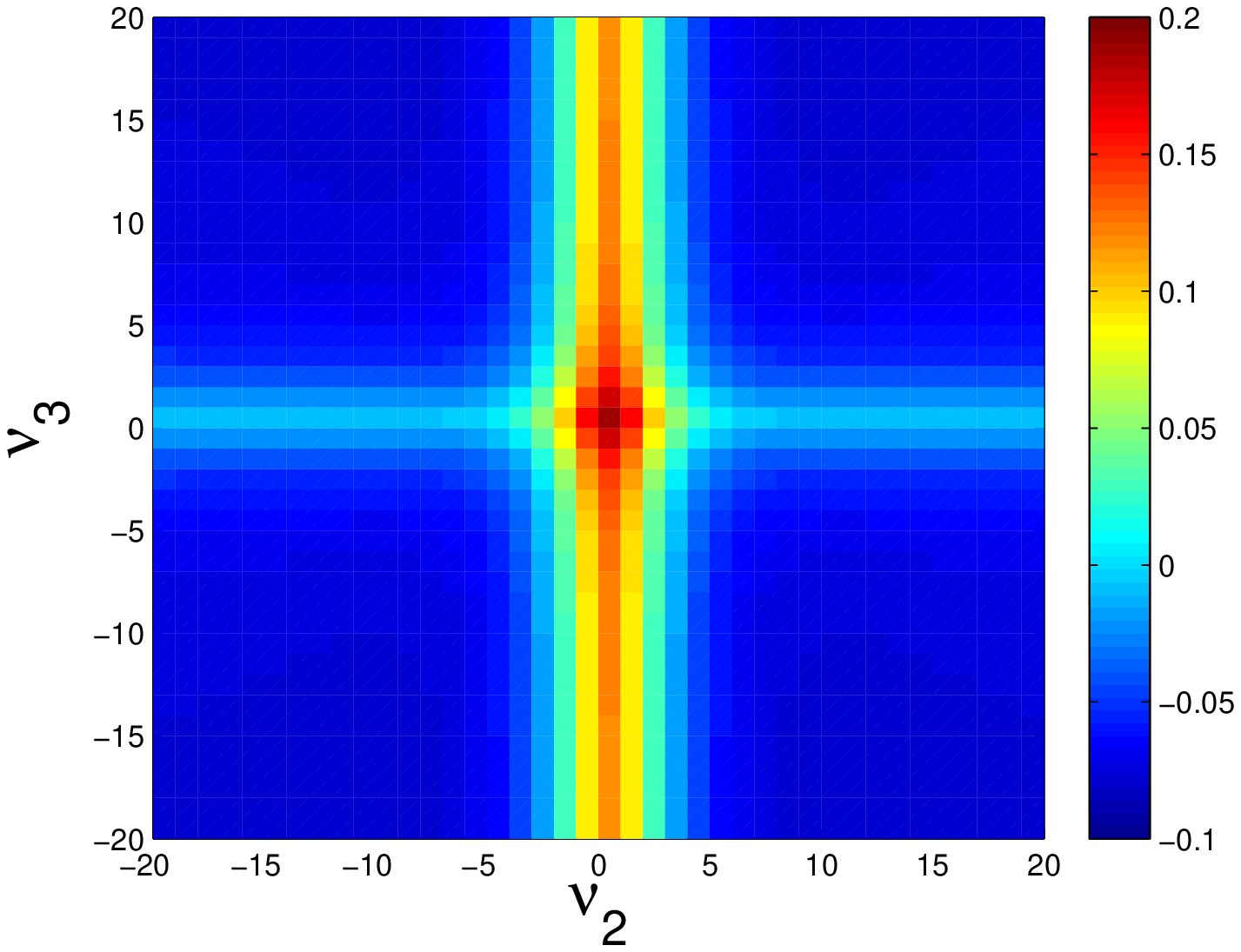}
    \includegraphics[width=0.4\textwidth]{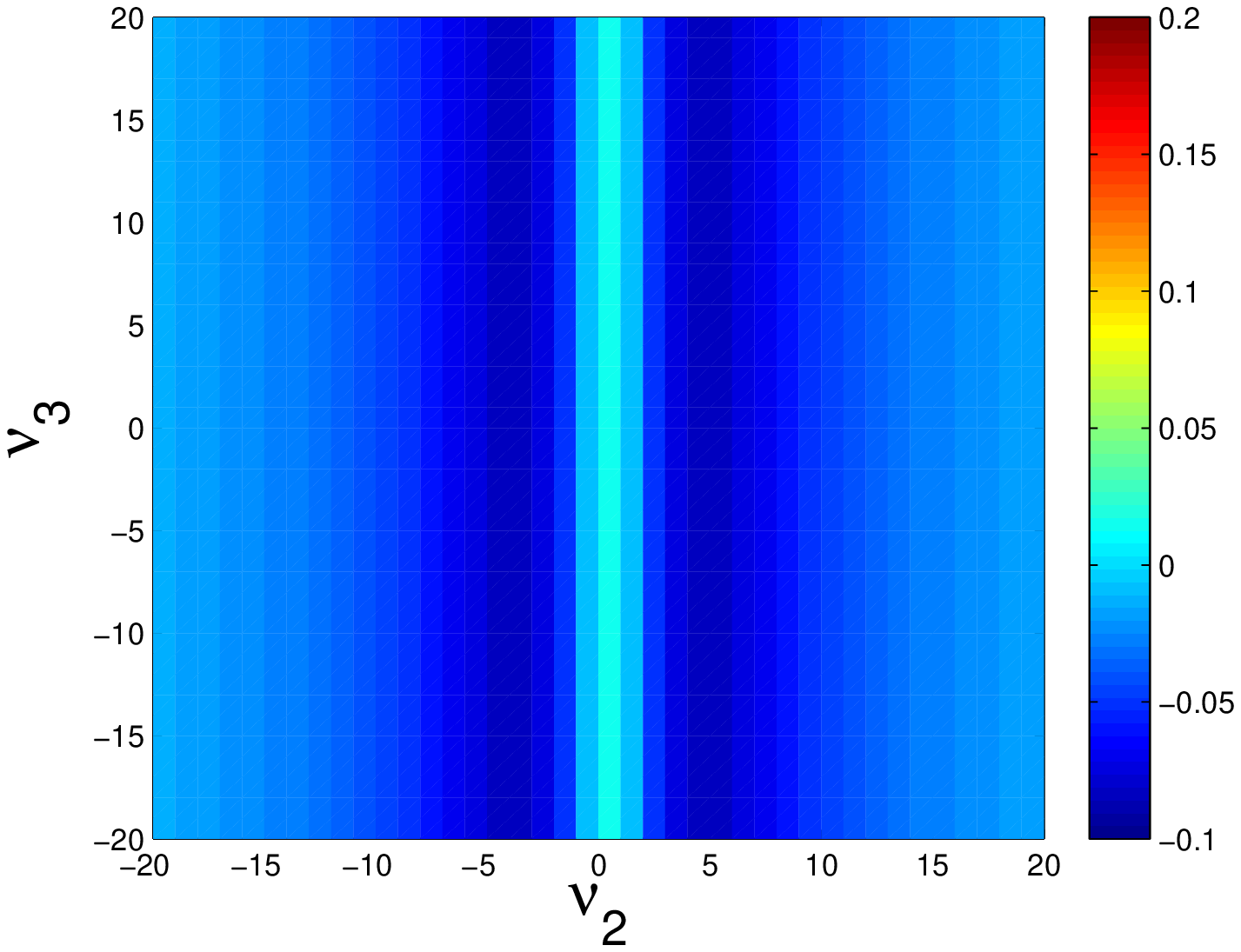}\\
    \includegraphics[width=0.4\textwidth]{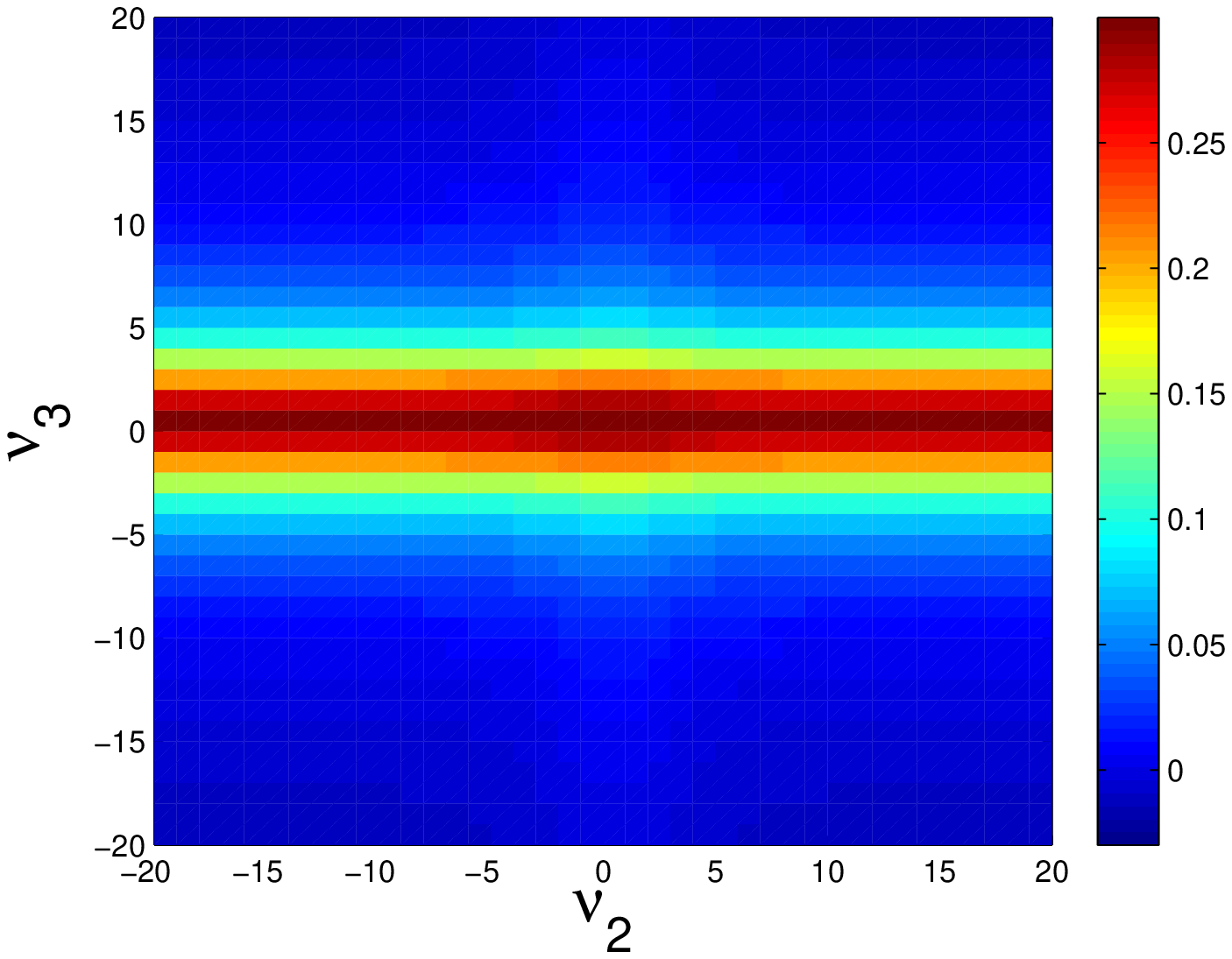}
    \includegraphics[width=0.4\textwidth]{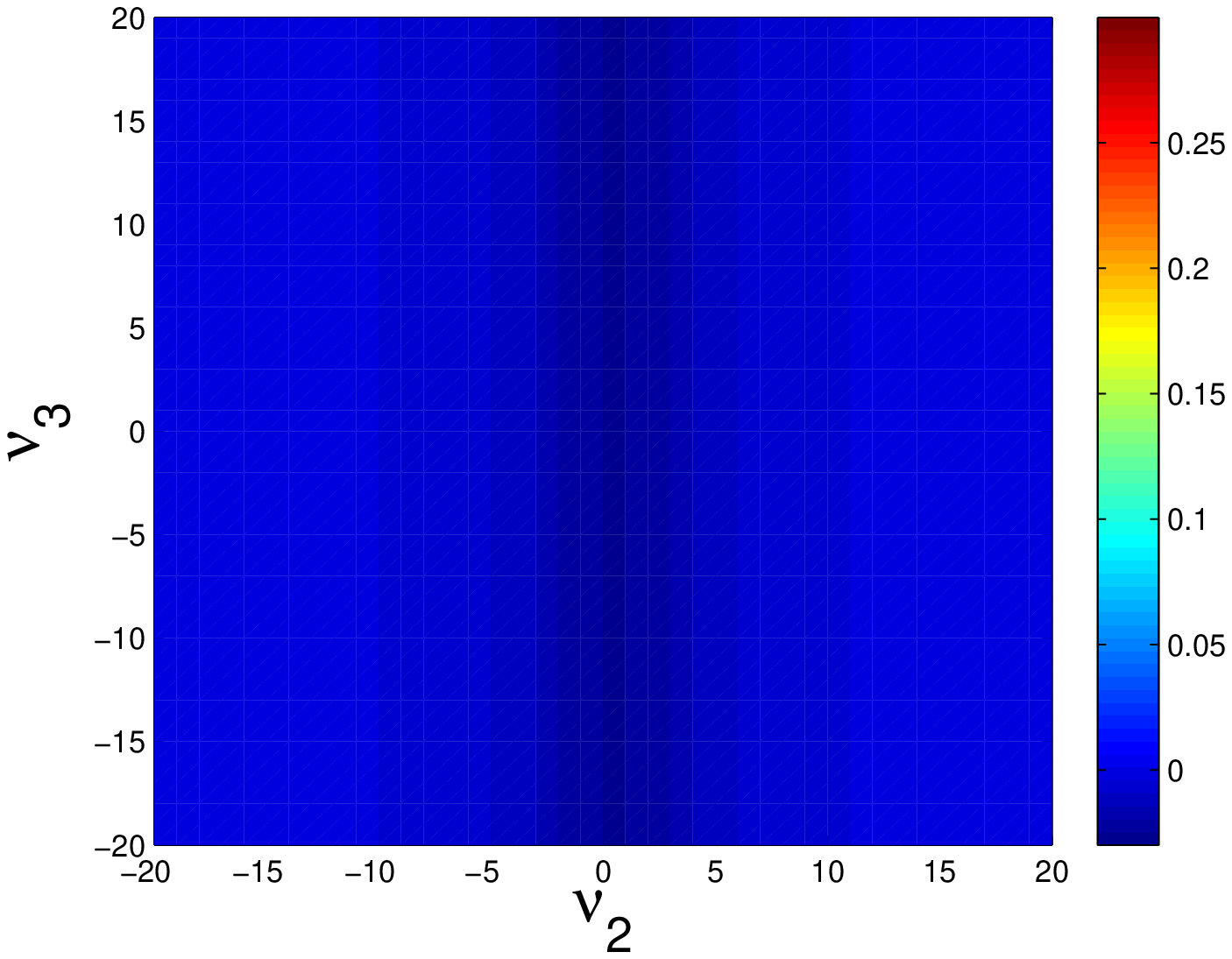}
    \caption[vertex]{(color online) Effective interactions at $U=3$. First row: The vertex $V^{\Lambda=1}(i\nu_1=0,i\nu_2,i\nu_3;1,1,1,1;d,d,d,d)-U$ in cfRG (left) and cRPA (right).
    Second row: The vertex $V^{\Lambda=1}(i\nu_1=0,i\nu_2,i\nu_3;1,2,1,2;d,d,d,d)-U$ in cfRG (left) and cRPA (right)
    Third row: The vertex $V^{\Lambda=1}(i\nu_1=0,i\nu_2,i\nu_3;1,2,2,1;d,d,d,d)-U$ in cfRG (left) and cRPA (right).}   
\label{pictvertexfrequencystructure}
\end{figure*}

In Fig.~\ref{pictvertexfrequencystructure} we show contour plots of the frequency-dependent part of the full vertex as sum of the three channels in the $i\nu_2$-$i\nu_3$-plane (with $i\nu_1=0$) for the cRPA and the cfRG calculation. Here the differences can be clearly read off. 
In the cRPA calculation the vertex only depends on transfer frequency $i\nu_2$, while in the cfRG calculation more frequency structure evolves, which leads to additional peaks at $i\nu_2$ (and also $i\nu_1$, not visible in the Figure).
These additional features are often comparable or larger than the structures that are found in the cRPA calculation.  In particular, the cfRG effective interactions can also enhance the bar interactions and cause 'anti screening', while the cRPA screening always tends to suppress the interactions. 

\subsection{Impact of new terms in effective Hamiltonian and interpretation}
We can now ask what the consequences of the additional interactions in the LEEM beyond the density-density terms are. One way to proceed with the full frequency structure would be to use CT-QMC to solve a small cluster (that could also be embedded via DMFT). As we do not have this heavy machinery available here, we resort to solving an approximate Hamiltonian that contains the new terms in exact diagonalization, again on a small number of $N$ sites corresponding to the wave vector discretization described before. 

\begin{figure}[htbp]
    \centering
    \includegraphics[width=0.450\textwidth]{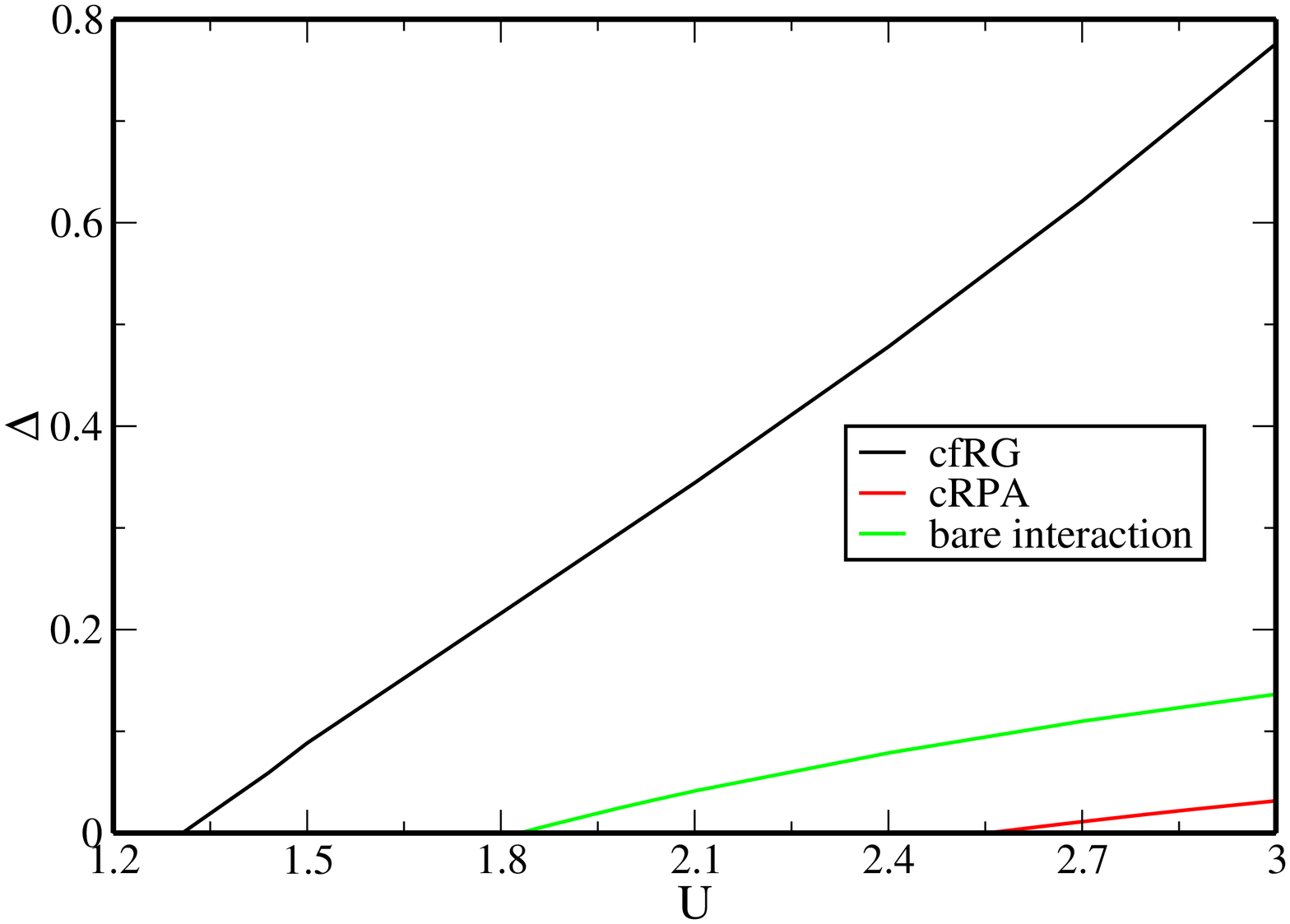}
    \includegraphics[width=0.450\textwidth]{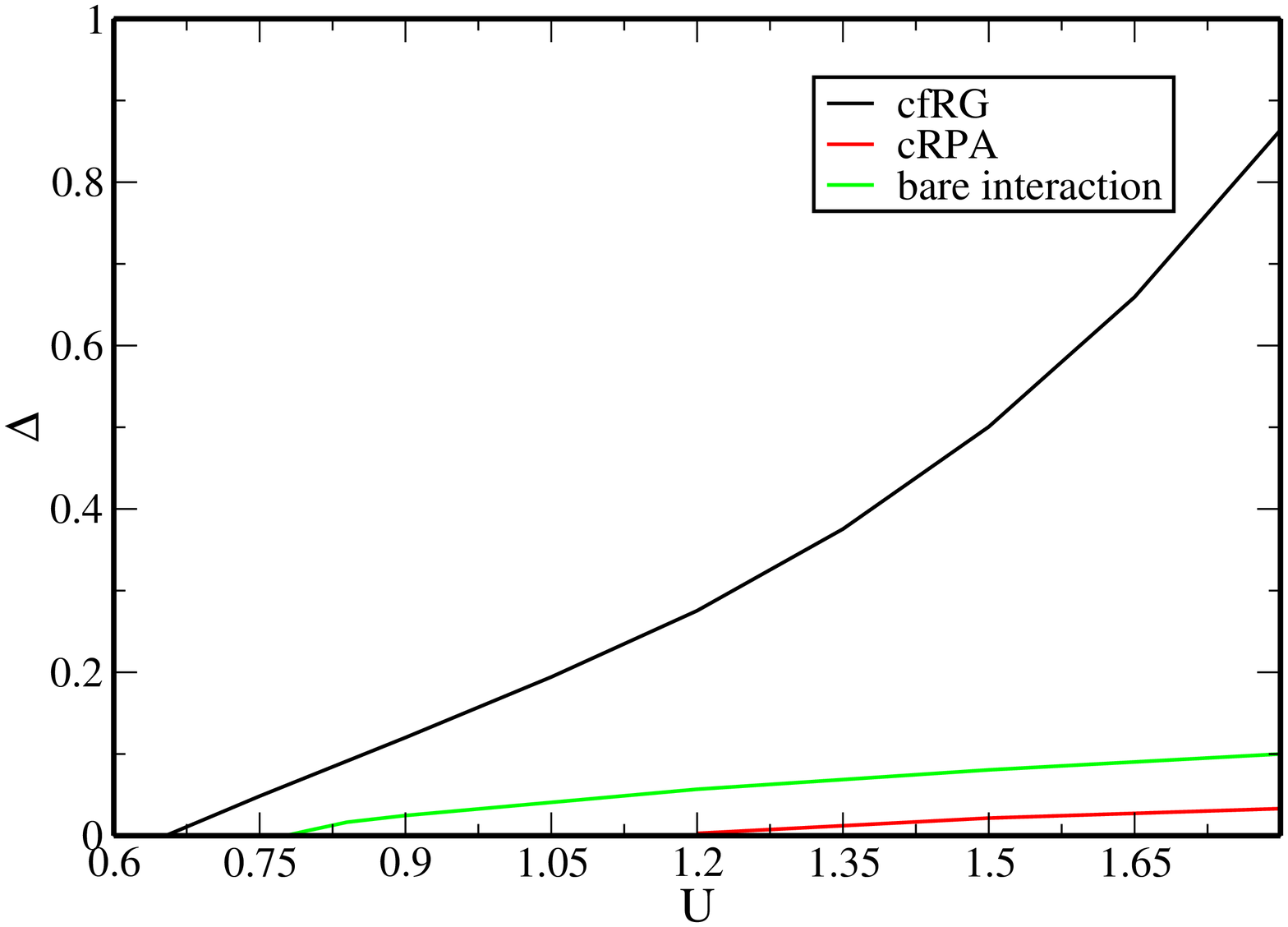}
    \caption[energieluecke]{(color online) Energy gap $\Delta$ between the high-spin groundstate and the first excited state as function of $U$ in scheme 1 (above) and
    scheme 2 (below) using three different approximations for the interaction terms in the exact solution of the effective $d$-band-model. In the simplest case we used just the bare interaction as input for the effective Hamiltonian. This is compared
    to the energy gap from the effective Hamiltonian with interaction parameters from the cRPA and the fRG calculation.} 
    \label{pictenergieluecke}
\end{figure}

In order to solve the effective theory for the $d$-like band we construct an effective Hamiltonian with the vertex functions at zero frequency transfer as interaction matrix elements, along the lines of Sec. \ref{vertex2hamilton}. 
More precisely we write
\begin{widetext}
\begin{equation}
\hat{H} =  Ê\sum_{k,\sigma}\epsilon_k d^{\dag}_{k,\sigma}d_{k,\sigma}+ \frac{1}{2} \sum_{\sigma,\sigma'}\sum_{i_{1'},i_{2'}\atop i_1,i_2}d^{\dag}_{i_{1'},\sigma}d^{\dag}_{i_{2'},\sigma'}V_{d}(i_{1'},i_{2'},i_1,i_2)d_{i_2,\sigma'}d_{i_1,\sigma}
\end{equation}
where the interaction constants are given by
\begin{equation}
\label{ }
V_d(i_{1'},i_{2'},i_1,i_2)=V^{\Lambda=1}(i\nu_1=0,i\nu_2=0,i\nu_3=0,i_{1'},i_{2'},i_1,i_2,d,d,d,d)
\end{equation}
\end{widetext}
These interaction terms can be interpretated in terms of charge, spin and pairing interactions according to Eqs. \ref{eqinterpretationdph},\ref{eqinterpretationcrph} and \ref{eqinterpretationpp}.
Instead of using the interaction terms at $i\nu_j=0$, we could also use $i\nu_j$-averages in each direction $j=\text{pp},\text{dph},\text{crph}$ which would lead to slightly smaller interaction terms. But on a qualitative level we do not expect differences in the results and the detailed choice of approximation will be of minor importance.
In this way all retardation effects are neglected, but we are able to solve the effective Hamiltonian exactly. In both schemes, i.e. for $N=2$ and for $N=4$,
we found that for a high enough interaction strength $U$ the groundstate at half filling of the remaining $d$-band is given by the respective highest spin state. In scheme 1 with 2 sites the groundstate spin is given by $S=1$
and in scheme 2 with 4 sites it is $S=2$. As a measure for the strength of the spin-spin interaction we used the 'spin  gap' $\Delta$ between the groundstate and the first
excited state (with spin $S=0$ in scheme 1 and $S=1$ in scheme 2). In Fig.~\ref{pictenergieluecke} we show $\Delta$ as function of $U$ in both schemes for three
different types of approximation. In the simplest 'tree-level' approximation we used only the bare interaction (i.e. without integrating out the excited energy levels) in the
effective Hamiltonian for the exact diagonalization. In the more advanced approximations we integrated out the excited levels by the cRPA and the fRG.

It turns out that the high-spin groundstate is already found in the tree-level approximation, where we use the bare interaction parameters for the effective model. So the tendency toward a high-spin state is not purely an effect from integrating out the high energy $s$- and $p$-like bands. Note however that this part of the interaction is frequency independent and cannot be interpreted straightforwardly as an interaction term like Eq.(\ref{eqspinspininteraction}).

Compared to this bare case the energy gap {\em decreases} in both discretization schemes ($N=2$ and $N=4$) when we integrate out the high energy levels in the cRPA scheme and {\em increases} 
in the cfRG scheme. In all cases there is a minimal threshold value of $U$ above which $\Delta$ becomes zero (and negative for smaller values of $U$). In scheme 1 with just two wave vector states this can be
easily understood, as there are only two energy levels in the effective Hamiltonian at $k=0$ and $k=\pi$ with different onsite-energies $\epsilon_{k=\pi}<\epsilon_{k=0}$.
For a small interaction strength both electrons occupy the $k=\pi$-level forming a spin singlet state. For a larger interaction strength the double occupancy becomes energetically unfavorable
and the electrons start to occupy the $k=0$-level. Due to a positive exchange interaction between both levels they form a high spin state. 

Very roughly, this increased tendency toward spin polarization can be understood as an inter band RKKY(Ruderman-Kittel-Kasuya-Yosida )-mechanism.\cite{Faz99} In our case the role of the conduction band that mediates the spin-spin interaction is taken by the high-energy bands, while the localized moments that interact are represent by the (still mobile) electrons in the LEEM target band.
In the usual RKKY effect for a metal, the spin susceptibility of the metal causes a power-law spin-spin interaction with a sign structure controlled by the Fermi surfaces geometry. Only for distances less than the Fermi wavelength, the mediated interaction is ferromagnetic, beyond this scale the sign of the interaction oscillates.  In our case, there are no particle-hole pairs that happen solely near the Fermi level, and the wave vector structure of the particle-hole bubbles is quite mild. Hence there are no sign oscillations, and at least within the range of our calculation, the interband-mediated spin-spin interaction remains ferromagnetic.

Note that we have here studied a weakly-correlated electron system and ignored some known effects. For example, a one-dimensional half-filled conduction band with repulsive interactions will become insulating due to Umklapp processes. This effect is not quite visible in our solution of the small cluster which is dominated by finite-size gaps.
For larger (for Cu more realistic) local  interaction parameters the Mott-insulating state at half filling will feature antiferromagnetic exchange interactions between the localized electrons. This mechanism will very likely outweigh the ferromagnetic tendencies described above. Yet, on a quantitative level, the corrections to cRPA should remain present.

\section{Conclusions}\label{conclusions}
We have studied the perturbative calculation of effective interaction parameters in multiband models with short-ranged interactions for electrons in solids. We have described how the functional renormalization group (fRG) can be employed to obtain these interaction terms of the low-energy effective model (LEEM) and how this treatment goes beyond the commonly used constrained random phase approximation (cRPA). For two models, we compared the cfRG  scheme and the cRPA. In the graphene-type model, both approximations gave similar results. The reason for the absence of noticeable corrections to the cRPA could be identified as the different symmetry with respect to $z$-inversion between the $\pi$-orbitals forming the conduction band and the $\sigma$-orbitals forming the bands that are integrated out. When the $z$-inversion is destroyed by additional terms in the free Hamiltonian, new terms beyond cRPA enter the cfRG flow and the LEEM interactions from the cfRG differed considerably form the cRPA. 
This opens a new way how effective interaction parameters of a system can be tuned, i.e. when the embedding of the system breaks a certain symmetry and hence makes non-cRPA contributions nonzero.
Rather generally, there are two other important channels beyond the cRPA in this setup: the particle-particle channel that in general suppresses the interactions further, and also the crossed particle-hole channel that is responsible for magnetic tendencies and that works to increase the effective interactions. Depending on the model parameters, not much of the original cRPA screening may be left in the cfRG effective interactions.

There are various paths that need to be pursued now:

1) The solution of the LEEM should be improved, already for the rather simplistic full models analyzed here. Then the meaning and impact of the terms beyond cRPA will become more clear. Besides employing state-of-the-art impurity solvers for clusters together with embedding schemes, nanosystems may represent a fruitful test arena, as additional benchmarking should be possible.

2) The formalism presented here needs to be cast in the more general framework of ab-initio calculations. Most likely, it will not be possible to capture all bands in the cfRG, as this would lead to an extreme numerical load, but ist is well conceivable that improved approximation scheme will treat the high-engery bands closest to the target bands in cfRG, while for higher-lying bands cRPA or simpler approximations are possible. Note also, that the inversion of the dielectric constant matrix that makes the cRPA potentially  a numerically complicated task is not necessary in the cfRG. Here the solution of the differential equation takes over this part.

3) We have focussed on the effective interaction of the low-energy model. Of course, self-energy effects might also be an important factor. In principle, the flow of the self-energy is also described by the fRG formalism, we have however not yet analyzed how this should be implemented in the question of deriving effective low-energy models. A recent study of  Shinaoka et al.\cite{Shi14} studies exactly this issue by powerful (extended) DMFT methods. The main upshot of their work is that a self consistent treatment of the energetics of the bands is essential, and that interacting propagators should be used if one tries to compute the effective interactions. It will be interesting to see how our work and the approach by Shinaoka et al. can be tied into a comprehensive picture.

4) In the present work, the initial interactions were taken to be short-ranged while in the ab-initio context Coulomb interactions should be considered. One of the reasons for this simplification is that any additional wave vector dependence, in particular one related to longer-ranged interactions, increases the numerical effort. In this paper, our main goal was to expose that the inclusion of the additional diagrams can cause substantial changes, and this is clearly visible already in the frequency structure we focussed on. We however expect that also the wave vector structure of the effective interactions will turn out much richer in cfRG than in cRPA. This can already be inferred from computing second-order corrections. Another thing than can be expected is that in the limit of small-wavector transfers between legs $1$ and $1'$, the deviations from cRPA will become smaller, as then the RPA diagrams of a given order will be helped by at least one factor $1/qa$ (with lattice constant $a$) compared to any other diagrams. Note however that one of the main applications for the effective interactions is to distill from them effective local and short-ranged interactions. Hence, one will typically average over the wave vector transfers and specialties at small transfers will not drastically influence the results. Therefore the deviations from cRPA for generic momentum transfers will not be negligible. \\

Acknowledgements: We thank S. Andergassen, S. Bl\"ugel, P. Hansmann, S.A. Maier, and T. Wehling  for helpful discussions. This project is supported by the DFG grants Ho2422/10-1 and Ho2422/11-1 (Austrian Science Fund SFB ViCoM) and by FOR 912. 

\bibliography{literatur}
\end{document}